%% file: composabilityTR.tex
\documentclass[11pt]{article}
\usepackage[margin=1in,letterpaper]{geometry}
\usepackage{amssymb,graphicx,graphics}
\usepackage{float}
\usepackage{color}
\usepackage{graphicx}
\usepackage{wrapfig}
\usepackage{psfrag}
\floatstyle{boxed}
\usepackage{hyperref}
\hypersetup{
 pdfborder = 0 0 0,
 colorlinks=true,
 citecolor=blue,
 backref=true,
}
\usepackage[hyperpageref]{backref}

\newtheorem{theorem}{Theorem}

\newtheorem{lemma}[theorem]{Lemma}
\newtheorem{definition}{Definition}

\newtheorem{prop}{Proposition}[definition]

\newcommand{\qed}{\hfill $\rule{1.7mm}{1.7mm}$}
\newcommand{\eat}[1]{}

\title{A New Look at Composition of Authenticated Byzantine Generals\\ }

 \author{Anuj Gupta\footnote{Department of Computer Science and Engineering, 
   Indian Institute of Technology Delhi, Hauz Khas, New Delhi-110016, India.}
   \\{\tt \small agupta@cse.iitd.ac.in} \and Prasant
   Gopal\footnote{Computer Science and Artificial Intelligence
     Laboratory(CSAIL), Massachusetts Institute of Technology,
     Cambridge, MA-02139, USA.}\\{\tt \small prasant@csail.mit.edu}\\ \\
   \and Piyush Bansal\footnote{Center for Security, Theory \& Algorithmic 
 	Research (CSTAR), International Institute of
     Information Technology, Hyderabad-500032, India.}\\{\tt \small
     piyush\_bansal@research.iiit.ac.in} \and 
     Kannan Srinathan$^\ddagger $ \\{\tt \small srinathan@iiit.ac.in} }

\date{}
\begin{document}

\maketitle
\thispagestyle{empty}

\begin{abstract}

  The problem of {\em Authenticated Byzantine Generals} (ABG) aims to
  simulate a {\em virtual}\/ reliable broadcast channel from the {\em
  General}\/ to all the players via a protocol over a {\em real}\/
  (point-to-point) network in the presence of faults. We propose 
  a new model to study the self-composition of ABG protocols. 
  The central dogma of
  our approach can be phrased as follows: Consider a player who
  diligently executes (only) the delegated protocol but the adversary
  steals some private information from him. Should such a player be
  considered faulty? With respect to ABG protocols, we argue that 
  the answer has to be {\em no}.\\

  In the new model we show that in spite of using unique session
  identifiers, if $n < 2t$, there cannot exist any ABG protocol that
  composes in parallel even twice. Further, for $n \geq 2t$, we design
  ABG protocols that compose for any number of parallel
  executions. Besides investigating the composition of ABG under a
  new light, our work also brings out several new insights into Canetti's
  Universal Composability framework.  Specifically, we show that
  there are several undesirable effects if one deviates from our
  dogma.  This provides further evidence as to why our dogma is the
  right framework to study the composition of ABG protocols.\\


\end{abstract}
\newpage
\setcounter{page}{1}
\section{Introduction}\label{intro}
The goal of an {\em Authenticated Byzantine Generals} (ABG) protocol is
to simulate a {\em virtual}\/ reliable broadcast channel from the {\em
  General}\/ to all the players via a protocol over a {\em real}\/
(point-to-point) network in the presence of faults. In general, one
may wish to simulate a virtual network which may not only involve some
reliable/secure channels but also reliable/secure nodes (like a
Trusted Third Party). Traditionally, the notion of faults in the
network is captured via a fictitious entity called {\em adversary}\/
that can choose to actively corrupt up to any $t$ of the $n$
players. A player is said to be non-faulty if he executes only the
delegated protocol code and does no more. In contrast, the computation
performed by the faulty players is chosen by the adversary.

It is well known that protocols which correctly simulate the desired
virtual network may not remain correct when run in presence of
same/other protocols~\cite{C01:UCS:aNPfCP}. For most real life
networks such as the Internet, a protocol is seldom executed in a stand
alone setting. Composition of protocols aims to study the
correctness/security of protocols when several protocols are run
concurrently.

Over the past few decades security of cryptographic protocols in stand
alone settings has been studied fairly well. Canetti extended these
studies to {\it arbitrary} unknown environments by introducing the
framework of \emph{Universal Composability} {\sf (UC)}.
Ever since, several researchers have built on it to prove many
exciting results. Besides providing a rigorous and an elegant
mathematical framework for proving security guarantees of protocols
that run in arbitrary environments, {\sf UC} is a classic notion that
has seemingly brought secure function evaluation and multi-party
computation ever close to practice.

\subsection{Prior Work}
Byzantine Generals Problem (BGP) and Byzantine 
Agreement (BA) were first introduced by Pease \emph{et al.}\/~\cite{LSP82,PSL80}. 
It is well known that BGP (likewise BA) over a
completely connected synchronous network is possible if and only if
$n>3t$~\cite{PSL80,LSP82}. Later on, the problem was studied in many
different settings,
giving both possibility (protocols) and impossibility
results. Some of the prominent settings are - incomplete networks~\cite{Dolev82}, 
probabilistic correctness~\cite{Rabi83}, asynchronous networks~\cite{FLP85}, 
partially synchronous networks~\cite{DolevDwStoc87}, mobile adversaries \cite{G94:RAitPoMF},
non-threshold adversarial model~\cite{fitzi98efficient},
mixed adversarial model~\cite{GP92,BerndFiziMaurer}, hypergraphs~\cite{FM00} to name a few.

Pease \emph{et al.}\/~\cite{PSL80,LSP82} introduced the
problem of \emph{authenticated Byzantine Generals} (ABG). 
Here, the players are augmented with 
Public Key Infrastructure(PKI) for digital signatures 
to authenticate themselves and their messages. Pease \emph{et al.}\/
proved that in such a model tolerability against a $t$-adversary can
be amazingly increased to $n>t$, which is a huge improvement over
$n>3t$. Being a reasonably realistic model and because of its high fault tolerance, 
ABG is an important and popular variant of BGP and hence, has been faily well studied. 
Dolev~\cite{dolev:ABA} proved that any ABG protocol 
over a completely connected synchronous network of $n$ nodes 
tolerating a $t$ Byzantine adversary will require $t+1$ rounds of communication. 
Further, he proposed algorithms that takes $O(t+1)$ rounds and $O(nt)$ messages.
Authenticating every message being sent by a player can be an expensive. Some 
works have explored cost cutting by considering alternatives to authentication
and limiting the use of signatures. Specifically, Borcherding~\cite{Borcherding:WDAG95,borcherding-partially}
explored the possibility of using signatures in only some rounds and not all. 
An alternative line of thought was suggested by Srikanth and
Toueg~\cite{SrikanthToueg:87} wherein authenticated messages are
simulated by non-authenticated sub-protocols. 
In another work, Borcherding ~\cite{Borderding96} 
studied different levels and styles of authentication and its effects on the agreement
protocols. This work focuses on understanding the properties of authentication scheme
their impact on building faster algorithms for BGP. 
Gong {\em et al.}\/~\cite{gong95byzantine}
study the assumptions needed for the authentication mechanism in
protocols for BGP that use signed messages. They propose protocols 
for BA that add authentication to oral message protocols so as
to obtain additional resilience due to authentication.
Schmid and Weiss~\cite{SW04:algs} study ABG under hybrid filure model of node and communication failures.
Katz \emph{et al.}\/~\cite{KatzKoo} propose expected constant 
round ABG protocols in the case $n>2t$. 
Gupta \emph{et al.}\/~\cite{GGBS10} study ABG in a model
where in the adversary can corrupt some players actively and some more players passively. 
Further, they require the passively corrupt player to be consistent with the honest players. 
They show that their model unifies the results of $n>3t$ (BPG) and $n>t$ (ABG).  
\cite{GJRA2010} too explores ABG in a partially compromised signature setting. 
Bansal \emph{et al.}\/~\cite{PPAS2011} extend the studies of \cite{GGBS10} 
to the case arbitrarily connected (undirected) networks.


Security of protocols under composition was first investigated in \cite{Ore,MR91,GO,GK}. 
Owing to its impact on the \emph{modular} approach of constructing cryptographic 
protocols, composition of protocols has been well studied in literature. 
Goldreich and Krawczyk~\cite{GK} studied sequential and parallel composition zero-knowledge
protocols. They proved that zero-knowledge and strong 
formulation of zero-knowledge (e.g. black box simulation) are not closed under parallel execution.
Richardson and Kilian~\cite{RK99} examined the concurrent composition of zero-knowledge
proofs. Canetti \emph{et al.}~\cite{CKPR} proved that Black-box concurrent zero-knowledge requires 
$\omega (\log n)$ rounds. Dwork \emph{et al.}~\cite{DNS} show that under the assumption of a restricted 
adversary (they call it ($\alpha$,$\beta$) constraint) there exists perfect concurrent 
zero-knowledge arguments for every language in class $NP$.
Canetti~\cite{canetti00security} proposed generic definitions 
of security for multi-party cryptographic protocols and proved that the security 
under these definitions continue to hold 
under the natural composition operation~\cite{MR91}. Canetti~\cite{C01:UCS:aNPfCP} 
introduced {\sf UC} to study the
security/correctness of protocols when run with arbitrary unknown
protocols. \cite{CF01:UCC,CK02:UCNoKEaSC} study commitments and key exchanges under composition.
Canetti \emph{et al.}~\cite{Canetti:2002:UCT} show that any two-party and multi-party functionality
is closed under universal composition, irrespective of the number of corrupted players. 
Canetti and Rabin~\cite{CR03:UCwJS} initiated the study of universal composition with joint state.
Ben-Or \emph{et al.}~\cite{benor05universal} took up the study of universal composability in quantum key distribution.
Some of the recent papers on protocol composition are~\cite{L03:CoSMPP:aCS,PS04:NNoS:AUCwTS,hofheinz09polynomial,CKS11,CJ11,Ostrovsky:2012:143}.

Lindell \emph{et al.}\/~\cite{LLR02:OtCoABA}  studied the properties of
self composition of ABG. They proved that, over a completely
connected synchronous network of $n$ players in presence of a
$t$-adversary, if $n \leq 3t$, then there \emph{does not} exist any ABG
protocol that self-composes in parallel even twice. Further, for $n>3t$, they designed
ABG protocols that self-compose in parallel for any number of
executions. Thus, proving the bound of $n>3t$ to be tight.  In the
same work, they also show that if one assumes additional facility of
\emph{unique session identifiers}, fault tolerance for ABG under
parallel self-composition can be restored back to $n > t$.

The work closest (yet, incomparable) to our line of thought is the
model considered by Canetti and Ostrovsky~\cite{Canetti99Honest}. They use a slightly different perspective as
to what a fault ``means''. In particular, they operate under a model
where in all the parties (even uncorrupted ones) may deviate from the
protocol but under the sole restriction that most parties do
not risk being detected by other parties as deviating from the
protocol.\\

\noindent{\bf Organization of the Paper.} In Section~\ref{section2},
 we present our case as to why the study of self composition of ABG protocols needs 
a new approach. This renders the model used in
the extant literature (to study the protocol composition) inappropriate for,
 at least, a few problems such as ABG and necessitates the formulation 
of a better model.  In Section~\ref{model}, we propose 
a new model to study protocol composition. We prove our results in 
Section~\ref{characterization}. 


\section{The Central Dogma} \label{section2}
Consider a player who is concurrently executing several protocols which
run as processes. Clearly, the player is faulty if any one of these
processes deviates from the originally designated protocol
code. However, there can be faulty players wherein some of the
processes continue to diligently execute the delegated protocol
code. {\em Should such processes be deemed faulty?} The answer can be
{\em no}\/ since they diligently execute the delegated protocol,
hence, they are certainly not Byzantine faulty. On the other hand, the
answer can be {\em yes}\/ because the player is faulty and therefore,
the {\em data private to such processes can always be accessed by the
adversar}y\footnote{Any process with administrative privileges can
always read the data internal to any other process within the same system.}.

All of literature on composition of protocols and in particular, 
the most recent one on ABG \cite{LLR02:OtCoABA} has considered 
such processes to be Byzantine faulty. However, 
in Section~\ref{subsec21} and~\ref{subsec22} we argue as to why it is better to model such
processes as passively corrupt (similar to honest but curious parties). 

\subsection{To Make Them Faulty or Not ?}
\label{subsec21}
The extant literature on (unauthenticated) reliable broadcast requires all non-faulty
players to agree on the same value~\cite{PSL80,LSP82}. 
Consequently, a player can be non-faulty in the following two ways -- (i)
The adversary is absent and (therefore) the player follows the
delegated protocol. (ii) The adversary is present, but allows the
player to diligently follow the delegated protocol and therefore, by the virtue of diligently following 
the protocol the player is non-faulty. 

With respect to ABG, the answer does not reveal itself automatically.
The issue with ABG is more subtle because
ABG is spawned by interests across various disciplines. In particular, 
ABG has continuously drawn inspiration from cryptography and in particular 
secure multi-party computation (SMPC).  So, in the case of ABG, any attempt to settle the question 
must consider the cryptographic viewpoint. When it comes to defining faults 
in SMPC, recall that, Ber-Or {\em et al.}\/~\cite{BGW88:CTfNFDC} 
define a player as \emph{faulty}\/ if and only if the player deviates from the designated
protocol. Therefore, w.r.t ABG, we have a choice -- 
(i) To punish the player by labelling him as faulty, if the adversary steals 
any private information (such as digital signature) from the player, 
despite the player diligently executing the designated protocol. 
(ii) To reward the player for diligently following the protocol and pay him back for his efforts by not 
labelling him as faulty (remember, that this player has certainly
helped in the simulation of broadcast channel by routing several crucial messages). 
In continuation, it is natural to brand all processes, whose private information is stolen by the adversary during a
cryptographic protocol, as faulty.  However, we believe w.r.t. ABG, the answer has to be the other way around.
One may argue that reliable broadcast is essentially a primitive in distributed computing
and that authentication was introduced only as a tool. 
Rather, authentication was only a means to facilitate the end (reliable broadcast).  
This is, however, hardly any reason.  To find the answer, one must journey to the very heart 
of every protocol for Byzantine Generals (BG).

The purpose of any BG (likewise ABG) protocol is to simulate a (virtual) reliable broadcast channel 
over a point-to-point network. Consider a scenario wherein the General is connected to all the players 
via an an actual physical broadcast channel. All the players including those under the adversary's control 
will always receive the same message from the General. The adversary can make the players under his 
influence to discard this message and deviate from the protocol. However, if the adversary chooses 
not to do so for any of the player(s) under his control, then such a player(s) will be in agreement 
with the group of players who were honest. Therefore, \emph{any ABG (BG) protocol aiming to truly simulate a
broadcast channel must ensure consistency between all the players who follow the designated protocol}.


\indent We now elaborate the implications of our dogma on the composition
of ABG protocols. It is well known that in the stand alone execution model, a
$t$-adversary is free to corrupt up to any $t$ players. With respect
to parallel composition of protocols, a $t$-adversary is free to choose any
set of $\leq t$ players and corrupt them in {\em all}\/ or {\em only some}\/ of
the executions. This permits the adversary to corrupt
different players in different executions i.e. a $t$-adversary {\em may} as
well corrupt, say $t_1$ players ($t_1 <t$) in some of the parallel
execution(s) and a different set of $t_2$ players ($t_2 <t$) in the remaining 
execution(s). As long as $t_1+t_2 \leq t$, w.r.t composition, such an
adversary is a valid $t$-adversary. 

\begin{figure}[h]
\begin{minipage}{3in}{{\hspace*{3cm}
\includegraphics[width=0.40\textwidth]{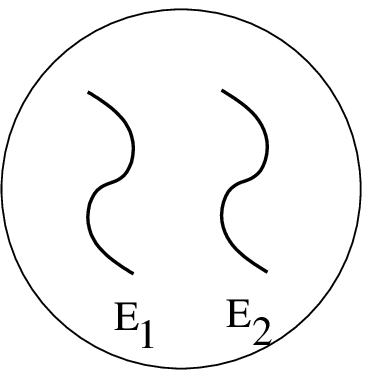}}}
\caption{\small {\it \label{2image1} $P$ running processes $E_1$ and $E_2$.}}
\end{minipage}\
\begin{minipage}{3in}{{\hspace*{3cm}
\includegraphics[width=0.40\textwidth]{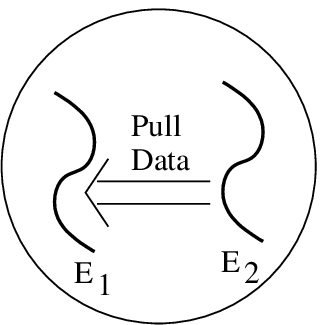}}}
\caption{\small {\it \label{2image2} By corrupting $P$ in $E_1$, adversary can always read the complete data of $P$ in $E_2$.}}
\end{minipage}\ 
\end{figure}

The above leads to an interesting observation w.r.t composition of ABG protocols --
by Byzantine corrupting a player in {\em some and not all} parallel executions, the
adversary can forge messages on behalf this player even in those
executions wherein this player is uncorrupted. We facilitate the same with the help 
of the following simple scenario: Consider a player $P$ running two parallel executions, 
say $E_1$ and $E_2$, of some (correct) ABG protocol (Figure~\ref{2image1}). 
Further, $P$ uses distinct authentication keys, say $k_1$ and $k_2$ in the
executions $E_1$ and $E_2$ respectively. The adversary corrupts $P$
in Byzantine fashion only in $E_1$. Consequently, the adversary can forge messages 
on behalf of $P$ in $E_2$ even though $P$ is non-faulty in $E_2$. 
\vspace{10pt}
\vspace{10pt}
This is because in $E_1$ the adversary can delegate that code to $P$
which can read the key $k_2$ (or for that
matter any private data) from the process $E_2$ (Figure~\ref{2image2}). The above observation 
can be extended to any number parallel executions $E_1, E_2, \ldots, E_k$. 
It is easy to see that the above observation holds good 
even if $P$ uses a {\em distinct} authentication
key in each of the parallel executions $E_1, E_2, \ldots, E_k$.

\subsection{New insights into current model}
\label{subsec22}
Besides providing a new model for composing ABG protocols, our work
also draws attention to a few (not so serious, yet interesting) grey
areas in the existing popular models of protocol composition.  These issues
deserve to be discussed in greater detail and we deem that these
aspects must be studied in depth before one can unleash the
paradigm of composability into the real world. The following can
also be viewed as the ill-effects of modelling the non-faulty
processes within a faulty player as Byzantine faulty:

\begin{enumerate}
\item {\em Proving incorrect protocols as correct:} A very popular
  paradigm used in the literature to capture the security/correctness
  requirements of a cryptographic protocol is the
  ideal-world/real-world simulation
  paradigm~\cite{C01:UCS:aNPfCP}. Informally, in this paradigm, a
  protocol is said to be correct if for every real world adversary
  ${\cal A}$ there exists an ideal world adversary ${\cal S}$ that
  can match the views of the players and the adversary in the two
  worlds. We now show that if the non-faulty processes within a faulty
  player are treated as Byzantine faulty, then, using the
  ideal-world/real-world simulation paradigm, the protocols which are
  incorrect in the stand alone settings \emph{can be} proven to be
  correct under composition!

Example: Consider the problem of secure addition -- players $\{P_a,
  P_b, P_c, P_d \}$ start with input values $\{v_a, v_b, v_c, v_d\}$
  respectively and wish to find the sum of their combined input values
  without revealing their input value to any other player. Consider a
  protocol $\xi$ which gives a random value as the answer. Clearly,
  in a stand alone setting $\xi$ is an incorrect protocol. However,\eat{using the
  Ideal-world/real-world simulation paradigm,} one can prove $\xi$ to be
  a correct protocol under composition as follows: Let $E_1$ and $E_2$ 
  be two concurrent executions of
  $\xi$. The real world adversary ${\cal A}$ corrupts
  player $P_a$ actively only in $E_1$. In $E_2$, $P_a$ executes
  the code exactly as per the specifications of $\xi$.  If $P_a$ is
  treated as Byzantine faulty in $E_2$, then in the ideal world
  execution the ideal world adversary ${\cal S}$ can actively corrupt
  $P_a$ in $E_2$ as well. Thus, in the ideal world execution of $E_2$,
  ${\cal S}$ sends that input value to the TTP(Trusted Third Party)
  which ensures that the view of the players and the adversary in the
  ideal world and real world is same. Specifically, if in $E_2$, the protocol
  $\xi$ gives the answer as $v_a+v_b+v_c+v_d+r$ where is $r$ is some
  positive random number. Then, in the corresponding ideal world
  execution, ${\cal S}$ on behalf of $P_a$ sends $v_a+r$ as the input
  value to the TTP. This will ensure that the views in the two worlds
  are same. 

\item {\em Basing security on internal communication:} Let $\Lambda$ be
  a secure composable protocol for some problem $\zeta$. From $\Lambda$
  one can always construct a new protocol $\Lambda^\prime$ as follows:
  $\Lambda^\prime$ is exactly same as $\Lambda$ except for the following two
  changes -- (i) Every process in $\Lambda^\prime$ sends all it's
  data (including its private data) to all the concurrent processes within
  the same player. (ii) Every process in $\Lambda^\prime$ ignores
  this incoming data from any of the fellow concurrent process within the same player. Clearly, if
  $\Lambda^\prime$ is secure then so is $\Lambda$. However, \emph{is $\Lambda^\prime$
  secure given $\Lambda$ is secure?}  The answer can be {\em no}\/ if, in $\Lambda^\prime$, the
  faulty processes chose {\em not} to ignore the incoming data.

  This implies that the security definition is dependent on the
  internal communication between the processes within a
  player. Clearly, one will prefer to have a security definition which
  does {\em not} depend on such intricate details. As highlighted by Canetti~\cite{canetti00security},
  this preference stems from the need and benefits of a simple, intuitive and workable security definition. 

\end{enumerate}

In essence, our dogma is the following: All non-faulty
processes, i.e. the processes within a non-faulty player that
execute the delegated protocol diligently and do no more, are considered
{\em honest}\/. All Byzantine faulty processes in a Byzantine faulty
player are considered {\em corrupt}\/. All non-faulty processes within a
Byzantine faulty player are considered {\em passively}\/ corrupt. We
aim to study self composition of ABG protocols in this new paradigm.

\section{Model}\label{model}
We are now ready to present our model. Our model is same as the
one used in the extant literature~\cite{C01:UCS:aNPfCP, LLR02:OtCoABA}
except for the following (small but important) changes -- (i) All non-faulty processes within a
non-faulty player are considered as {\em honest}\/. (ii) All Byzantine
faulty processes within a Byzantine faulty player are considered as {\em corrupt}\/. 
(iii) All non-faulty processes within a Byzantine faulty
player are treated as {\em passively}\/ corrupt. ((iii) follows from the
observations made in Section~\ref{section2}). Here, a process is said to be non-faulty if it
exactly executes the delegated protocol code and does no more. Further,
a process is said to be Byzantine faulty if it executes the
program code of adversary's choice. 

We consider a set of $n$ players, $\mathbb{P}$=$\{p_1,p_2, \ldots, p_n\}$,
over a completely connected synchronous network. Any protocol in this
setting is executed in a sequence of {\em rounds}\/ where in each
round, a player can perform some local computation, send new messages
to all the players, receive messages sent to him by other players in
the same round, (and if necessary perform some more local
computation), in that order. The notion of faults in the system is
captured by a virtual entity called \emph{adversary}. During the
execution, the (polynomial-time) adversary\footnote{Digital signatures
  based authentication necessitates the assumption of a
  polynomial-time adversary. Our impossibility proofs do not need this
  assumption but our protocols require a ``magical" means to
  authenticate if the adversary is unbounded.} may take control of up
to any $t$ players and make them behave in any arbitrary fashion. Such
an adversary is called as a $t$-adversary. Further, the players can
invoke multiple parallel executions of any protocol. We model this via
players running multiple processes in parallel. We assume that
the communication channel between any two players is perfectly
reliable and authenticated. We also assume existence of a
(signature/authentication) scheme via which players authenticate
themselves. This is modelled by all the players having an additional
setup-tape that is generated during the preprocessing phase. Note that
keys cannot be generated with in the system itself. Similar to ~\cite{LLR02:OtCoABA}, 
it is assumed that the keys are generated using a trusted system 
and distributed to players prior to running of the protocol. Typically, in such a preprocessing phase,
signatures and verification keys are generated. That is, each player
gets his own private signature key, and in addition, public
verification keys for all the other players. No player can forge any
other player's signature and the receiver can uniquely identify the
sender of the message using the signature. However, the adversary can
forge the signature of all the $t$ players under its control. The
adversary can inject forged messages, on behalf passively corrupt
processes, via Byzantine corrupt processes. Further, we assume that
each run of a protocol is augmented with \emph{unique session identifiers} (USIDs).

\subsection{Defining Composable ABG}

We use the well established ideal/real process simulation paradigm
to define the requirements of ABG. Both the ideal process and the
 real process have the set ${\mathbb P}$ of $n$ players including 
the General ${\cal G}$ as common participants. Apart from these, the 
ideal process has a TTP (Trusted Third Party) and an ideal process
 adversary ${\cal S}$ whereas the real process has a real process 
adversary ${\cal A}$. We start by defining the ideal process for ABG.\\

\noindent \textbf{Ideal process ($\Psi_{ideal}$):} 
(1) ${\cal G}$ sends his value $v$ to TTP. (2) TTP forwards the same 
to all the $n$ players and $\cal{S}$. (3) All honest players output 
$v$. $\cal{S}$ determines the output of faulty players.\\
We assume that all message transmissions in the above protocol are perfectly secure. \\

\noindent Let $IDEAL_{TTP,\cal{S}}(v,r_{\cal{S}},\overrightarrow{r})$
denote a vector of outputs of all $n$ players running $\Psi_{ideal}$
where ${\cal G}$ has input $v$, ${\cal S}$ has random coins $r_{\cal{S}}$
and $\overrightarrow{r}$ where $\overrightarrow{r}$= $r_1, r_2, \ldots, r_n,r_{TTP}$;
$r_1, r_2, \ldots, r_n$ and $r_{TTP}$ are the random
coins of $n$ players and the TTP respectively.
$IDEAL_{TTP,\cal{S}}(v)$ denotes the random variable describing
$IDEAL_{TTP,\cal{S}}(v,r_{\cal{S}},\overrightarrow{r})$ when $r_{\cal{S}}$ 
and $\overrightarrow{r}$ are chosen uniformly at random. $IDEAL_{TTP,\cal{S}}$ 
denotes the ensemble $\{IDEAL_{TTP,\cal{S}}(v)\}_{v \in {\{0,1\}}}$. \\

\noindent \textbf{Real life process ($\Psi_{real} (\Pi)$):} Unlike 
in the ideal process, here the players interact among themselves 
as per a designated protocol $\Pi$ and the real process adversary 
${\cal A}$. Specifically: (1) Every honest player proceeds according 
to the protocol code delegated to him as per $\Pi$. (2) The adversary 
${\cal A}$ may send some arbitrary messages (perhaps posing as any 
of the corrupt players) to some/all of the players. (3) Honest 
players output a value as per $\Pi$. ${\cal A}$ determines the 
output of faulty players.\\

\noindent Let $REAL_{\Pi,\cal{A}}(v,r_{\cal{A}}, \overrightarrow{r})$
denote a vector of output of all $n$ players running
$\Psi_{real}(\Pi)$ where ${\cal G}$ has input $v$, and
$r_{\cal{A}}$,$\overrightarrow{r} = r_1, r_2, \ldots, r_n$ are the
random coins of the adversary and $n$ players respectively.  Let
$REAL_{\Pi,\cal{A}}(v)$ denote the random variable describing
$REAL_{\Pi,\cal{A}}(v,r_{\cal{A}},\overrightarrow{r})$ when
$r_{\cal{A}}$ and $\overrightarrow{r}$ are chosen uniformly at
random. Let $REAL_{\Pi,\cal{A}}$ denote the ensemble
$\{REAL_{\Pi,\cal{A}}(v)\}_{v \in {\{0,1\}}}$.\\

\noindent We directly adopt the definitions of Lindell {\em et al.}\cite{LLR02:OtCoABA}.

\begin{definition}[ABG]\label{defABG}
  $\Pi$ is an ABG protocol tolerating a
  $t$-adversary if for any subset $I \subset {\mathbb P}$ of
  cardinality up to $t$ (that is , $|I| \leq t$), it holds that for
  every probabilistic polynomial-time real process adversary ${\cal A}$
  that corrupts the players in $I$ in $\Psi_{real} (\Pi)$, there
  exists a probabilistic polynomial-time ideal process adversary
  $\cal{S}$ in $\Psi_{ideal}$ that corrupts the players in $I$, such
  that the ensembles $IDEAL_{TTP,\cal{S}}$ and $REAL_{\Pi,\cal{A}}$
  are computationally indistinguishable.
\end{definition}

\begin{definition}[Composable ABG~\cite{LLR02:OtCoABA}]\label{defABGCompose}
  Let $\Pi$ be an ABG protocol. $\Pi$ is said to remain secure 
  under parallel composition if for every polynomial time 
  adversary ${\cal A}$, the requirements for ABG
  (which is elaborated in Definition~\ref{defABG}) hold for $\Pi$ for
  every execution within the following process: Repeat the following
  process in parallel until the adversary halts:
\begin{enumerate}
\item The adversary ${\cal A}$ chooses the input $v$ for the General ${\cal G}$.
\item All players are invoked for an execution of $\Pi$ (using the
  strings generated in the preprocessing phase and an {\em unique
    session identifier}\/ for this execution). All the messages sent
  by the corrupted players are determined by the adversary ${\cal A}$,
  whereas all other players follow the instructions of $\Pi$.
\end{enumerate}
\end{definition}

Furthermore, as noted by Lindell {\em et al.}, Definition~\ref{defABGCompose} 
implies stateless composition i.e. all honest players are
oblivious to the other executions taking place in
parallel. In contrast, the adversary ${\cal A}$ can coordinate between
the parallel executions, and the adversary's view at any given time
includes all the messages received in all the executions.

\subsection{Our Results}\label{Results}
Recall that in the absence of unique session identifiers, 
ABG is not self-composable even twice if $n \leq 3t$~\cite{LLR02:OtCoABA}. 
We prove that unique session identifiers aid in improving the fault-tolerance 
of ABG protocols (that compose in parallel) but from $n>3t$ only to $n \geq 2t$. 
We, now, present the main theorem of this paper:

\begin{theorem}[Main Theorem]\label{OurResultsThm}
 ABG over $n$ players, tolerating a $t$-adversary, can be self-composed 
 in parallel for any number of executions if and only if $n \geq 2t$.
\end{theorem}

To put things in perspective, one can achieve the bound of $n>t$ for (a 
simplified variant of) ABG~\cite{LLR02:OtCoABA} if one makes the following 
assumption: {\em Only those non-faulty processes that run in non-faulty 
players need to be consistent, others need not, where a process is faulty if it 
deviates from the designated protocol.} 

\section{Complete Characterization}\label{characterization}

The aim of this section is to prove aforementioned Theorem~\ref{OurResultsThm}. We begin with a few definitions:

\begin{definition}[Adversary Structure]
An adversary structure ${\cal Z}$ for the player set 
${\mathbb P}$ is a collection of plausible sets of 
players which can be corrupted by the adversary. 
Formally, ${\cal Z} \subseteq 2^{\mathbb P}$, 
where all subsets of $Z$ are in ${\cal Z}$ if $Z \in {\cal Z}$. 
\end{definition}

\begin{definition}[Adversary Basis]
For an adversary structure ${\cal Z}$, $\bar{\cal Z}$ 
denotes the basis of the structure, i.e. the set of 
the maximal sets in ${\cal Z}$: $\bar{\cal Z} = 
\{Z \in \bar{\cal Z}: \nexists Z^\prime \in \bar{\cal Z}: Z \subset Z^\prime \}$
\end{definition}

\subsection{Qualifiers}\label{Starters}
We first prove that there \emph{does not} 
exist any ABG protocol that self-composes in parallel even 
twice over a network of 3 players, $\mathbb{P}$=\{$A$,$B$,$C$\}, tolerating an adversary basis $\bar{\cal
  A}=\{((C),(A)); ((A),(\emptyset));$\\$((B),(A))\}$. Here, (($x$),($y$))
represents a single element of the adversary basis such that the adversary
can Byzantine corrupt $x$ and $y$ in the first and second parallel execution, respectively. 
For the rest of this paper $\Pi_{k}$ (likewise $\Delta_{k}$) 
denotes an ABG protocol $\Pi$ ($\Delta$) that remains correct upto $k$ parallel self-compositions.

Before presenting the proof, we make a few comments on the proof style. As was 
with Lindell {\em et al.} (the base case of their work draws inspiration from \cite{FLM85}),
we establish ours on \cite{GGBS10}.  We, however, note that the overlap ends there. We remark that this
is not a serious concern and if at all everything but a testimony to the impact of \cite{FLM85}. \\

\begin{wrapfigure}{r}{0.30\textwidth}
\vspace{-35pt}
\begin{center}
\input{1.pstex_t}
\end{center}
\vspace{-20pt}
\caption{Network ${\cal N}$.}
\label{5fig1}
\vspace{-10pt}
\end{wrapfigure}
\vspace{-20pt}


\begin{theorem}\label{5theorem_motiv}
  There does not exist any 
  $\Pi_{2}$ over a network of 3 nodes, $\mathbb{P}$=\{$A$,$B$,$C$\}, tolerating an adversary basis 
  $\bar{\cal A}=\{((C),(A)); ((A),(\emptyset)); ((B),(A))\}$.
\end{theorem}
\emph{Proof:} Our proof demonstrates that the real process adversary (characterized by $\bar{\cal A}$)
can make the non-faulty processes in one of the parallel executions of any ABG protocol
to have an inconsistent output. In contrast, in the corresponding ideal world execution 
the non-faulty processes are guaranteed to have a consistent
output. It then follows that there \emph{does not} exist any ideal
world adversary ${\cal S}$ that can ensure that the output
distributions are similar. This violates Definition~\ref{defABGCompose}, hence the theorem.

To prove that $\bar{\cal A}$ can ensure that the non-faulty processes 
in one of the parallel executions do not have a consistent
output, we assume otherwise and arrive at a contradiction. We accomplish the same   
using the ideas from the proof technique developed by Fischer {\em et al.}\/~\cite{FLM85}. 

Formally, assume for contradiction that there exists a protocol $\Pi_{2}$ over
${\cal N}$ (Figure~\ref{5fig1}), $\mathbb{P}$=\{$A$,$B$,$C$\}, tolerating the adversary 
basis $\bar{\cal A}=\{((C),(A));((A),(\emptyset)); ((B),(A))\}$. 
Using $\Pi_{2}$, we create a protocol $\Pi^\prime$ [Definition~\ref{PiDash}] 
in such a way that existence of $\Pi_{2}$ implies existence 
of $\Pi^\prime$ (Proposition~\ref{PiPiD}). We, then, combine two copies of $\Pi^\prime$ to
construct a system ${\cal L}$ (Figure~\ref{5fig2}) and show that
${\cal L}$ must exhibit contradictory behaviour. It then follows that
the assumed protocol $\Pi_{2}$ cannot exist. 

\begin{figure}[htbp]
\begin{center}
\input{2.pstex_t}
\caption{Construction of ${\cal L}$ using two copies of $\Pi^\prime$.}
\label{5fig2}
\end{center}
\vspace{-15pt}
\end{figure}

We do not know what system ${\cal L}$ solves. 
Formally, ${\cal L}$ is a synchronous
system with a well defined behaviour. That is, for any particular input assignment 
${\cal L}$ exhibits some well defined output distribution. We obtain a
contradiction by showing that for a particular input 
assignment, no such well defined behaviour is possible. 
No player in ${\cal L}$ is aware of the complete
system, rather each player is aware of only his immediate neighbours. In
reality, a player may be connected to either $A$ or $A^{\prime}$~\footnote{$A$ and 
$A^\prime$ are independent copies of $A$ with the same authentication key.}, but he cannot
distinguish between the two. He knows his neighbours only by their
local name, in this case, $A$. Further, all players in $\cal L$ are oblivious to the fact 
that there are duplicate copies of the nodes in ${\cal L}$.
Specifically, for all $X \in \{A,B,C\}$, ${\cal L}$ is constructed in a manner 
such that the in-neighbourhood of any node $X$(or $X^\prime$) in ${\cal L}$ 
is same as the in-neighbourhood of the corresponding node $X$ in
${\cal N}$. 

Let the players in ${\cal L}$ start with input 
values as indicated in Figure~\ref{5fig2}; and $\alpha$ be the resulting 
execution. All the players in
$\alpha$ are honest and diligently follow $\Pi^\prime$. 
Further, let $E_1$ and $E_2$ be two parallel executions of
$\Pi_{2}$ over ${\cal N}$. We, now, define three distinct scenarios -- $\alpha_1$, $\alpha_2$ and
$\alpha_3$ :
\begin{itemize}
\item {\bf $\alpha_1$:} In $E_1$, $A$ is the \emph{General} and starts
with input value $0$. $\bar{\cal{A}}$ Byzantine corrupts $C$ in $E_1$. 
In $E_2$ $\bar{\cal{A}}$ Byzantine corrupts $A$.
\item {\bf $\alpha_2$:} In $E_1$, $A$ is the \emph{General}. Further, in $E_1$, $\bar{\cal{A}}$
  corrupts $A$, interacts with $B$ as if $A$ started with input value $0$ 
  and interacts with $C$ as if $A$ started with input value $1$.
\item {\bf $\alpha_3$:} In $E_1$, $A$ is the \emph{General} and starts
with input value $1$. $\bar{\cal{A}}$ Byzantine corrupts
$B$ in $E_1$. In $E_2$ $\bar{\cal{A}}$ Byzantine corrupts $A$.
\end{itemize}

\begin{figure}[htbp]
\begin{center}
\input{3.pstex_t}
\caption{Player's output in execution $\alpha$. Contradiction is shown via red rectangle}
\label{5fig3}
\end{center}
\vspace{-10pt}
\end{figure}
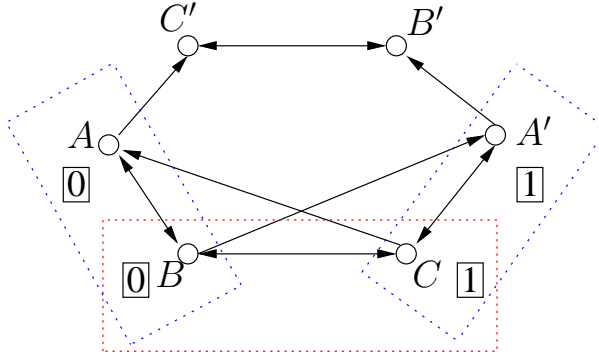

We claim that, in execution $\alpha$, both $A$ and $B$ decide 
on $0$. Similarly, we claim that both $C$ and $A^\prime$ in $\alpha$  
decide on $1$. Further, we claim that in $\alpha$ both $B$ and $C$ 
must decide on the same output (Figure~\ref{5fig3}). [To avoid clutter we defer the proof of these claims 
to Lemma~\ref{lbc3},~\ref{lbc5} and~\ref{lbc7}.]
However, in $\alpha$, players $B$ and 
$C$ have already decided on $0$ and $1$, respectively. Hence,
${\cal L}$ exhibits a contradictory behaviour. \qed \\

To complete the above proof, we now define the protocol 
$\Pi^\prime$ [Definition~\ref{PiDash}] and show that existence 
of $\Pi_{2}$ implies existence of $\Pi^\prime$ (Proposition~\ref{PiPiD}). 
We then prove the left out claims via Lemma~\ref{lbc2} - Lemma~\ref{lbc7}.

\begin{definition}[$\Pi^\prime$]\label{PiDash}
  For player $B$, every statement in $\Pi_{2}$ of
  the kind ``$B$ sends message {m} to $A$" is replaced by ``$B$ multicasts
  message {m} to all instances of $A$"(i.e. $A$, $A^\prime$) in
  $\Pi^\prime$. Similarly, every statement of the kind ``$C$ sends message
  {m} to $A$" in $\Pi_{2}$ is replaced by ``$C$ multicasts message {m} to
  all instances of $A$"(i.e. $A$, $A^\prime$) in $\Pi^\prime$. Rest all statements in
  $\Pi^\prime$ are exactly same as those in $\Pi_{2}$.
\end{definition}


\begin{prop}\label{PiPiD}
If $\Pi_{2}$ exists, then so does $\Pi^\prime$.
\end{prop}
\emph{Proof}: Follows directly from Definition~\ref{PiDash}. Given $\Pi_{2}$, one 
can always construct $\Pi^\prime$ by making appropriate changes in $\Pi_{2}$ as per the 
definition of $\Pi^\prime$. \qed \\

Rest of this section focuses on proving  Lemma~\ref{lbc2} - Lemma~\ref{lbc7}. 
The proofs are conceptually simple however, owing to the 
``topology" of system $\cal L$ and presence of authenticated messages, 
some of the proofs (Lemma~\ref{lbc2},~\ref{lbc4} and \ref{lbc6}) have tedious details.
A reader interested in the proof of our main theorem can directly jump to Section~\ref{MainCourse}.



We begin by introducing the terminology used in the proofs. Let $msg_{i}^{\Omega}(x,y)_x$ 
denote the message sent by player $x$ to player $y$ in round $i$ of execution $\Omega$. 
The $x$ in the subscript refers to the last player who authenticated this message. W.l.o.g, we assume 
that every player always authenticates every message sent by him. Further, let $V_{x,i}^{\Omega}$ 
denote {\em view} of player $x$ at the end of round $i$ in execution $\Omega$. Intuitively, 
$V_{x,i}^{\Omega}$ consists of everything that player $x$ ever ``sees" from round 1 until 
the end of round $i$ in execution $\Omega$. For our setting this includes (w.r.t execution $\Omega$) -- 
(i) Input value (if any) of $x$ : ${\cal I}_{x}^{\Omega}$. (ii) Secret key 
used by $x$ for authentication : ${\cal SK}_{x}^{\Omega}$. (iii) Protocol code executed by 
$x$ : $\theta_{x}^{\Omega}$. (iv) Set of all the messages sent by $x$ until the end of round 
$i$ : $\forall z \in  \mathbb{P}, ~\forall k \in (1,i), ~\bigcup_{k}(msg_{k}^{\Omega}(x,z)_x)$. 
(v) Set of all the messages received by $x$ until the end of round $i$ : 
$\forall z \in  \mathbb{P}, ~\forall k \in (1,i), ~\bigcup_{k}(msg_{k}^{\Omega}(z,x)_z)$. Formally:
\begin{eqnarray}\label{eqn1}
V_{x,i}^{\Omega} = \biggl[{\cal I}_{x}^{\Omega}, ~{\cal SK}_{x}^{\Omega}, ~\theta_{x}^{\Omega}, ~\bigcup_{k}(msg_{k}^{\Omega}(x,z)_x), ~\bigcup_{k}(msg_{k}^{\Omega}(z,x)_z) \biggr];~\forall z \in  \mathbb{P}, ~\forall k \in (1,i)
\end{eqnarray}

\noindent Since the messages sent by player $x$ in round $i$ of $\Omega$ is a function 
of $x$'s view until the end of round $i-1$ (i.e. $V_{x,i-1}^{\Omega}$), Equation~\ref{eqn1} 
can be rewritten as: 
\begin{eqnarray}\label{eqn2}
V_{x,i}^{\Omega} = ~ \biggl[{\cal I}_{x}^{\Omega}, ~{\cal SK}_{x}^{\Omega}, ~\theta_{x}^{\Omega}, ~\bigcup_{k}(msg_{k}^{\Omega}(z,x)_z) \biggr];~\forall z \in  \mathbb{P}, ~\forall k \in (1,i)
\end{eqnarray}

\noindent Our proofs will often have statements of the following form -- view of player $x$ until round $i$ of execution $\gamma$ is same as the view of player $y$ until round $i$ of execution $\delta$ (dubbed as $V_{x,i}^{\gamma} \sim V_{y,i}^{\delta}$). 
In order to prove such statements we will use the following observation:-

\noindent $V_{x,i}^{\gamma} \sim V_{y,i}^{\delta}$\/ if and only in the following conditions\footnote{~\cite{FLM85} referred to these conditions as \textbf{Locality Axiom}. In the case of ABG, the secret key used by a player is definitely a part of his view. Hence, we added condition (ii).} hold:\vspace{5pt} \\ 
~(i) ${\cal I}_{x}^{\gamma} = {\cal I}_{y}^{\delta}$\\
~(ii) ${\cal SK}_{x}^{\gamma} = {\cal SK}_{y}^{\delta}$\\
~(iii) ${\cal \theta}_{x}^{\gamma} = {\cal \theta}_{y}^{\delta}$\/~\footnote{We remark that 
${\cal \theta}_{x}^{\gamma}$ can differ from ${\cal \theta}_{y}^{\delta}$. However, as long 
as long as they generate the same message for any player $z$ i.e. $\forall x, y, z  
~msg_{i}^{\gamma}(x,z)_x = msg_{i}^{\theta}(y,z)_y$, it suffices for our proof.}\\
~(iv) $\forall z \in  \mathbb{P}, \forall k \in (1,i), ~msg_{k}^{\gamma}(z, x)_z = msg_{k}^{\delta}(z, y)_z$\\


\noindent Through out our proofs conditions (i), (ii) and (iii) will be trivially satisfied. Thus, our proofs will focus 
on proving condition (iv). For brevity we say:
\begin{eqnarray}\label{eqn3}
V_{x,i}^{\gamma} \sim V_{y,i}^{\delta}\/ \mbox{ iff }\mbox{ } \forall z \in  \mathbb{P}, \mbox{ } \forall k \in (1,i), \mbox{ } msg_{k}^{\gamma}(z, x)_z = msg_{k}^{\delta}(z, y)_z
\end{eqnarray}


\noindent Let $V_{x}^{\gamma}$ denote view of $x$ at the end of execution $\gamma$. Then, 
\begin{eqnarray}\label{eqn4}
V_{x}^{\gamma} \sim V_{y}^{\delta}\/ \mbox{ iff }\mbox{ } \forall k>0, \mbox{ } V_{x,k}^{\gamma} \sim V_{y,k}^{\delta}
\end{eqnarray}


\noindent Combining ($3$) and ($4$), we get
\begin{eqnarray}\label{eqn5}
V_{x}^{\gamma} \sim V_{y}^{\delta}\/ \mbox{ iff }\mbox{ } \forall z \in  \mathbb{P}, \mbox{ } \forall k >0, \mbox{ } msg_{k}^{\gamma}(z, x)_z = msg_{k}^{\delta}(z, y)_z
\end{eqnarray}

\begin{figure}[htbp]
\begin{center}
\input{induction_sub_trees_11.pstex_t}
\caption{$T_{A}^{\alpha}$ and $T_{A}^{E_1:\alpha_1}$ at the end of $k$ rounds.}
\label{exec_trees}
\end{center}
\end{figure}
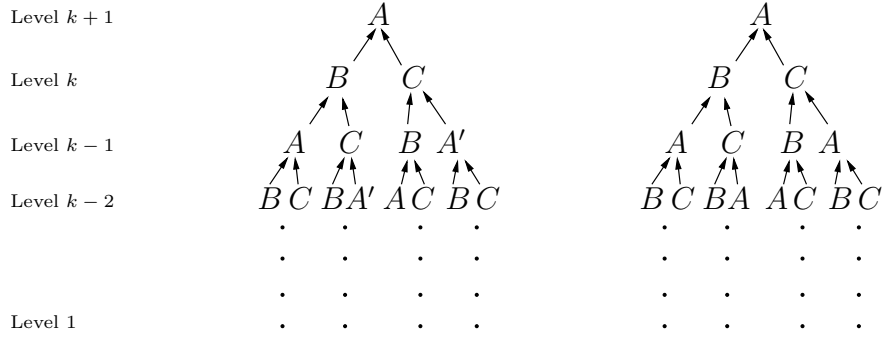

Though conceptually simple, proving the right hand side of Equation~\ref{eqn5} can be a tedious task. 
This is because the use of authentication limits the adversary's ability to send forged messages. 
Hence, one must formally establish that the adversary can indeed ensure that the right hand side 
of Equation~\ref{eqn5} holds true. To facilitate the same, we introduce 
the notion of {\em execution trees}. To understand the utility of execution trees, 
let us revisit the scenarios $\alpha$ and $\alpha_1$ as defined in the proof of Theorem~\ref{5theorem_motiv}. 
Recall that the proof requires us to show that $\forall z \in  \{A,B,C\}, \mbox{ } \forall k > 0, 
\mbox{ } msg_{k}^{\alpha}(z, A)_z = msg_{k}^{E_1:\alpha_1}(z, A)_z$. Note that what $A$ receives 
in round $i$ of $\alpha$ (likewise $E_1:\alpha_1$) depends on what $B$ and $C$ 
send him in round $i$ of $\alpha$ (likewise $E_1:\alpha_1$). So, we need to argue that these messages, sent 
in round $i$ of $\alpha$ and $E_1:\alpha_1$ respectively, are either same or {\em can be} made same by the adversary.
The messages sent by $B$ and $C$ in round $i$ of $\alpha$ (likewise $E_1:\alpha_1$) depend on what they 
themselves receive in round $i-1$. This in turn depends on what $A$ and $C$ (likewise, $A$ and $B$) send to 
$B$ ($C$) in round $i-2$ of $\alpha$ ($E_1:\alpha_1$). Thus, we need to argue 
that the adversary can ensure that whatever messages $A,C$ (likewise, $A$ and $B$) send to $B$ ($C$) in 
round $i-2$ of $\alpha$ is same as whatever messages $A,C$ ($A$ and $B$) send to $B$ ($C$) in 
round $i-2$ of $E_1:\alpha_1$. Note that this continues in a recursive manner until the recursion 
stops at round 1. The entire recursion can be visualized as trees, $T_{A}^{\alpha}$ and 
$T_{A}^{E_1:\alpha_1}$, rooted at $A$ for executions $\alpha$ and $E_1:\alpha_1$ 
respectively, as shown in Figure~\ref{exec_trees}.\\

\vspace{-10pt}

An {\em execution tree} can be regarded as a ``visual" analogue of Equation~\ref{eqn5}.
Formally, $T_{X}^{\Omega}$ is n-ary tree i.e. a node can have upto $n$ children. 
where $n$ is the number of players ($\mathbb P$) participating in execution $\Omega$. 
Each node has a label $l \in {\mathbb P}$. The root node of $T_{X}^{\Omega}$ 
has label $X$. The levels of the tree are named in a bottom up manner. 
The lower most level is 1, the one immediately above it is 2 and so on. 
A node $x$ is a child of node $y$ if and only if $x$ is in the in-neighbourhood 
of $y$ in execution $\Omega$. Thus the number of children of any node $z$ is same 
as the size of the in-neighbourhood of $z$. An edge from node 
$y$ at level $j$ to node $x$ at level $j+1$ in the tree represents the message 
that $y$ sends to $x$ in round $j$ of $\Omega$. All the edges in the tree are 
directed from child to parent and are between adjacent levels only. Let $T_{x,i}^{\Omega}$ 
denote the execution tree of $x$ until round $i$ in execution $\Omega$.

{\em We note the following} -- to show that a player, say $x$, receives same messages in two different executions, say $\gamma$ and $\delta$, it suffices to show that execution trees $T_{x}^{\gamma}$ and $T_{x}^{\delta}$ are similar. To prove this similarity, we will use induction on the heights of $T_{x}^{\gamma}$ and $T_{x}^{\delta}$. \\

\noindent As a prelude to proving Lemma~\ref{lbc2}, we present adversary's strategy in $E_1:\alpha_{1}$. Recall that we defined scenario $\alpha_1$ as: In $E_1$ $A$ is the \emph{General} and starts with input value $0$. $\bar{\cal{A}}$ Byzantine corrupts 
$C$ in $E_1$. In $E_2$ $\bar{\cal{A}}$ Byzantine corrupts $A$. \\

\noindent Adversary's ($\bar{\cal{A}}$) strategy in $E_1:\alpha_{1}$ is as follows:
\begin{enumerate}
\item \emph{Send outgoing messages of round i:} Based on the messages received during round $i-1$, $\bar{\cal{A}}$ decides on the messages to be sent in round $i$. For round 1, $\bar{\cal{A}}$ sends to $B$ what an honest $C$ would have sent to $B$ in execution $E_1:\alpha_2$. For $i\geq 2$, $\bar{\cal{A}}$ authenticates $msg_{i-1}^{E_1:\alpha_{1}}(B,C)_{B}$ using $C$'s key and sends it to $A$. For $msg_{i-1}^{E_1:\alpha_{1}}(A,C)_{A}$, $\bar{\cal{A}}$ examines the message. If the message has not been authenticated by $B$ even once, it implies that the message has not yet been seen by $B$. Then, $\bar{\cal{A}}$ sends a message to $B$ which is same as what $C$ would have sent to $B$ in round $i$ of execution $E_1:\alpha_{2}$. Formally, $\bar{\cal{A}}$ constructs $msg_{i-1}^{E_1:\alpha_{1}}(A,C)_{A}$ such that $msg_{i-1}^{E_1:\alpha_{1}}(A,C)_{A} = msg_{i-1}^{E_1:\alpha_{2}}(A,C)_{A}$. Note that this is possible because $A$ is actively corrupt in $E_2:\alpha_{1}$ and therefore $\bar{\cal{A}}$ can forge messages on behalf of $A$ in $E_1:\alpha_{1}$. $\bar{\cal{A}}$ then authenticates $msg_{i-1}^{E_1:\alpha_{1}}(A,C)_{A}$ using $C$'s key and sends it to $B$. However, if the message has been authenticated by $B$ even once, then $\bar{\cal{A}}$ simply authenticates $msg_{i-1}^{E_1:\alpha_{1}}(A,C)_{A}$ using $C$'s key and sends it to $B$.

\item \emph{Receive incoming messages of round i:} $\bar{\cal{A}}$ obtains messages $msg_{i}^{E_1:\alpha_{1}}(A,C)_{A}$ and $msg_{i}^{E_1:\alpha_{1}}(B,C)_{B}$ via $C$. (These messages are sent by $A$ and $B$ respectively to $C$ in round $i$). Similarly via $A$, $\bar{\cal{A}}$ obtains messages $msg_{i}^{E_1:\alpha_{1}}(B,A)_{B}$ and $msg_{i}^{E_1:\alpha_{1}}(C,A)_{C}$. (These are also round $i$ messages sent by $B$ and $C$ respectively to $A$. Players respectively compute these messages according to their respective view until round $i-1$).
\end{enumerate}


\begin{lemma}\label{lbc2}
$\bar{\cal A}$ can ensure $V_{A}^{\alpha} \sim V_{A}^{E_1:\alpha_{1}}$ and $V_{B}^{\alpha} \sim V_{B}^{E_1:\alpha_{1}}$.
\end{lemma}
\emph{Proof}: Using induction on $i$, we show that for any round $i$, $T_{A,i}^{\alpha} \sim T_{A,i}^{E_1:\alpha_{1}}$. It then follows that $T_{A}^{\alpha} \sim T_{A}^{E_1:\alpha_{1}}$. Combining this with 
Equation~\ref{eqn5} gives $V_{A}^{\alpha} \sim V_{A}^{E_1:\alpha_{1}}$. 

Note that owing to the topology of ${\cal L}$, only nodes present in $T_{A}^{\alpha}$ are $A$, $B$, $C$ and 
$A^\prime$. $B^\prime$ and $C^\prime$ do not occur in $T_{A}^{\alpha}$. Hence, $A^\prime$ has an outgoing 
directed edge only and only to $C$. Likewise, $A$ has an outgoing directed edge only to $B$. Corresponding 
nodes present in $T_{A}^{E_1:\alpha_{1}}$ are $A$, $B$, $C$ and $A$ respectively. We analyse 
$T_{A}^{\alpha}$ and $T_{A}^{E_1:\alpha_{1}}$ in a bottom up manner.

\begin{figure}[htbp]
\begin{center}
\input{subtree_0_1_1.pstex_t}
\caption{ $T_{A,1}^{\alpha}$ and $T_{A,1}^{E_1:\alpha_{1}}$.}
\label{trees_0_1_1}
\end{center}
\end{figure}
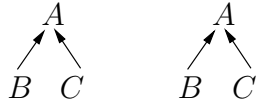

\vspace{5pt}
\noindent {\tt Base Case:} \\
\underline{$i=1$}\\
Consider round 1 of executions $\alpha$ and $E_1:\alpha_{1}$. Corresponding execution trees $T_{A,1}^{\alpha}$ and $T_{A,1}^{E_1:\alpha_{1}}$ are shown in Figure~\ref{trees_0_1_1}. Now, $B$ starts with same input, secret key and executes same code in $\alpha$ and $E_1:\alpha_{1}$ i.e ${\cal I}_{B}^{\alpha} = {\cal I}_{B}^{E_1:\alpha_{1}}$, ${\cal SK}_{B}^{\alpha} = {\cal SK}_{B}^{E_1:\alpha_{1}}$ and ${\cal \theta}_{B}^{\alpha} = {\cal \theta}_{B}^{E_1:\alpha_{1}}$ (the last equality follows from our definition of $\Pi^\prime$ [Definition~\ref{PiDash}] ). Thus, $B$ will send same messages to $A$ in round 1 of $\alpha$ and $E_1:\alpha_{1}$ i.e. $msg_{1}^{\alpha}(B,A)_{B} = msg_{1}^{E_1:\alpha_{1}}(B,A)_{B}$. Since $C$ is Byzantine corrupt in $E_1:\alpha_{1}$, $\bar{\cal A}$ can ensure that $msg_{1}^{\alpha}(C,A)_{C} = msg_{1}^{E_1:\alpha_{1}}(C,A)_{C}$. Thus, $T_{A,1}^{\alpha} \sim T_{A,1}^{E_1:\alpha_{1}}$.

\vspace{-10pt}

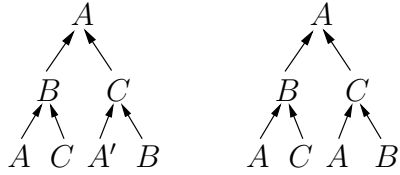
\begin{figure}[htbp]
\begin{center}
\input{subtree_0_1_2.pstex_t}
\caption{ $T_{A,2}^{\alpha}$ and $T_{A,2}^{E_1:\alpha_{1}}$.}
\label{trees_0_1_2}
\end{center}
\end{figure}

\noindent \underline{$i=2$}\\
We show that the similarity holds for round 2 as well. Consider $T_{A,2}^{\alpha}$ and $T_{A,2}^{E_1:\alpha_{1}}$ as shown in Figure~\ref{trees_0_1_2}. Now, ${\cal I}_{A}^{\alpha} = {\cal I}_{A}^{E_1:\alpha_{1}}$, ${\cal SK}_{A}^{\alpha} = {\cal SK}_{A}^{E_1:\alpha_{1}}$, ${\cal \theta}_{A}^{\alpha} = {\cal \theta}_{A}^{E_1:\alpha_{1}}$ and ${\cal I}_{B}^{\alpha} = {\cal I}_{B}^{E_1:\alpha_{1}}$, ${\cal SK}_{B}^{\alpha} = {\cal SK}_{B}^{E_1:\alpha_{1}}$, ${\cal \theta}_{B}^{\alpha} = {\cal \theta}_{B}^{E_1:\alpha_{1}}$. Thus, $msg_{1}^{\alpha}(A,B)_{A} = msg_{1}^{E_1:\alpha_{1}}(A,B)_{A}$ and $msg_{1}^{\alpha}(B,C)_{B} = msg_{1}^{E_1:\alpha_{1}}(B,C)_{B}$. Since $C$ is Byzantine corrupt in $E_1:\alpha_1$, $\bar{\cal A}$ can ensure that $msg_{1}^{\alpha}(C,B)_{C} = msg_{1}^{E_1:\alpha_{1}}(C,B)_{C}$. 

\vspace{5pt}
\vspace{-10pt}

Now, ${\cal I}_{A^\prime}^{\alpha} \neq {\cal I}_{A}^{E_1:\alpha_{1}}$, thus $msg_{1}^{\alpha}(A^\prime,C)_{A^\prime} \neq msg_{1}^{E_1:\alpha_{1}}(A,C)_{A}$. However, since $A$ is Byzantine corrupt in $E_2:\alpha_{1}$, $\bar{\cal{A}}$ can forge messages on behalf of $A$ in $E_1:\alpha_{1}$ (follows from observations made in Section~\ref{subsec21}). $\bar{\cal{A}}$ can use this to 
simulate $C$ having received same messages in round 1 in $E_1:\alpha_1$ and $\alpha$.

Specifically, as $C$ is Byzantine faulty in $E_1:\alpha_{1}$, $\bar{\cal{A}}$ can construct $msg_{1}^{E_1:\alpha_1}(A,C)_{A}$ in a way such that $msg_{1}^{E_1:\alpha_1}(A,C)_{A} = msg_{1}^{\alpha}(A^\prime,C)_{A^\prime}$. Now $B$ receives same round 1 messages in $E_1:\alpha_{1}$ and $\alpha$ and $B$ has same input value, secret key and executes same code, thus $msg_{2}^{E_1:\alpha_{1}}(B,A)_{B} = msg_{2}^{\alpha}(B,A)_{B}$. Thus, edge $BA$ (between levels $2$ and $3$) in $T_{A,2}^{E_1:\alpha_{1}}$ is same as the corresponding edge $BA$ in  $T_{A,2}^{\alpha}$. Since, $C$ is Byzantine corrupt $\bar{\cal{A}}$ can ensure that $msg_{2}^{E_1:\alpha_{1}}(C,A)_{C} = msg_{2}^{\alpha}(C,A)_{C}$. Therefore, $\bar{\cal{A}}$ can ensure that the edge $CA$ (between levels $2$ and $3$) in $T_{A,2}^{E_1:\alpha_{1}}$ is same as the corresponding edge $CA$ in  $T_{A,2}^{\alpha}$. Thus, $T_{A,2}^{E_1:\alpha_{1}} \sim T_{A,2}^{\alpha}$. \\

\noindent {\tt Induction hypothesis:} Let it be true for all rounds upto $k$ i.e. $\forall i, ~i \leq k$, $T_{A,i}^{\alpha} \sim T_{A,i}^{E_{1}:\alpha_{1}}$. Likewise $\forall i, ~i \leq k$, $T_{B,i}^{\alpha} \sim T_{B,i}^{E_{1}:\alpha_{1}}$\\

\noindent {\tt Induction step:} We now prove that the similarity holds for round $k+1$ as well i.e. $T_{A,k+1}^{\alpha} \sim T_{A,k+1}^{E_{1}:\alpha_{1}}$. 

\vspace{5pt}
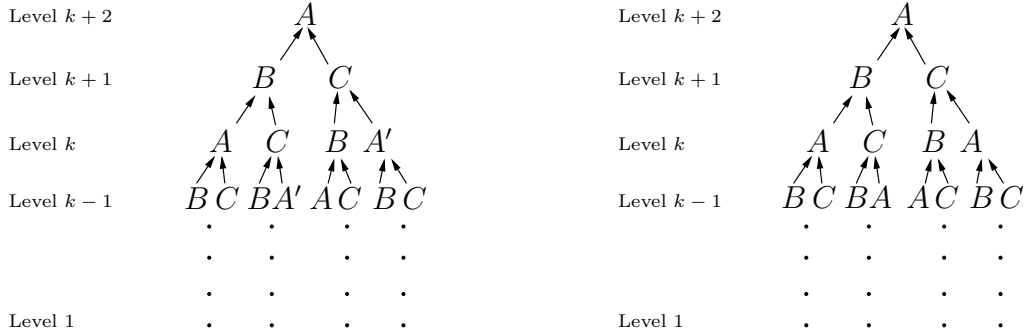
\begin{figure}[htbp]
\begin{center}
\input{induction_sub_tree_1.pstex_t}
\caption{$T_{A,k+1}^{\alpha}$ and $T_{A,k+1}^{E_1:\alpha_{1}}$.}
\label{induction_sub_tree_1}
\end{center}
\end{figure}

Consider $T_{A,k+1}^{\alpha}$ and $T_{A,k+1}^{E_1:\alpha_{1}}$ 
as shown in Figure~\ref{induction_sub_tree_1}. Consider the edges between level 
$k$ and $k+1$. From the induction hypothesis, we have $\forall j\leq k$, $T_{A,j}^{\alpha} 
\sim T_{A,j}^{E_1:\alpha_{1}}$. Further, ${\cal I}_{A}^{\alpha} = {\cal I}_{A}^{E_1:\alpha_{1}}$, 
${\cal SK}_{A}^{\alpha} = {\cal SK}_{A}^{E_1:\alpha_{1}}$, ${\cal \theta}_{A}^{\alpha} = {\cal \theta}_{A}^{E_1:\alpha_{1}}$.
Thus, $A$ sends same messages to $B$ in round $k$ of both the executions i.e. 
$msg_{k}^{E_1:\alpha_{1}}(A,B)_{A} = msg_{k}^{\alpha}(A,B)_{A}$. 
Thus, edge $AB$ (between levels $k$ and $k+1$) is same in both the trees. 
Likewise, from the induction hypothesis we have $\forall j\leq k$, $T_{B,j}^{\alpha} 
\sim T_{B,j}^{E_1:\alpha_{1}}$. Therefore, 
$msg_{k}^{E_1:\alpha_{1}}(B,C)_{B} = msg_{k}^{\alpha}(B,C)_{B}$. Hence, 
edge $BC$ (between levels $k$ and $k+1$) is same in both the trees.

Now, consider the nodes - $A^\prime$ at level $k$ in $T_{A}^{\alpha}$ and 
the corresponding node $A$ in $T_{A}^{E_1:\alpha_1}$. For time being assume\footnote{
Using induction on the value of $j$, one can show that the assumption is true.} 
that $\forall j\leq k$, $T_{A^\prime,j}^{\alpha} \sim T_{A,j}^{E_1:\alpha_{1}}$. We claim that 
$\bar{\cal A}$ can simulate $C$ at level $k+1$ in $T_{A}^{E_1:\alpha_1}$ to have received 
messages from $A^\prime$ exactly same as the messages received 
by $C$ at level $k+1$ in $T_{A}^{\alpha}$. This is because $A$ is 
Byzantine corrupt in $E_2:\alpha_{1}$, thus $\bar{\cal A}$ can forge messages 
on behalf of $A$ in $E_1:\alpha_{1}$. Formally, $\bar{\cal A}$ constructs 
$msg_{k}^{E_1:\alpha_1}(A^\prime,C)_{A^\prime}$ such that 
$msg_{k}^{E_1:\alpha_1}(A^\prime,C)_{A^\prime} = msg_{k}^{\alpha}(A,C)_{A}$.
Thus, $\bar{\cal A}$ can ensure that the edge $A^\prime C$ (one between 
levels $k$ and $k+1$) in $T_{A,k+1}^{E_1:\alpha_1}$ is same as the corresponding 
edge $AC$ in $T_{A,k+1}^{E_1:\alpha_1}$.

Now, ${\cal I}_{C}^{\alpha} = {\cal I}_{C}^{E_1:\alpha_{1}}$, 
${\cal SK}_{C}^{\alpha} = {\cal SK}_{C}^{E_1:\alpha_{1}}$, 
${\cal \theta}_{C}^{\alpha} = {\cal \theta}_{C}^{E_1:\alpha_{1}}$
Thus, $C$ sends same round $k+1$ messages 
to $A$ in $\alpha$ and $E_1:\alpha_{1}$ i.e. $msg_{k+1}^{E_1:\alpha_1}(C,A)_{C} = msg_{k+1}^{\alpha}(C,A)_{C}$. 
Thus, edge $CA$ (one between levels $k+1$ and $k+2$) in $T_{A,k+1}^{E_1:\alpha_1}$ 
is same as the corresponding edge $AC$ in $T_{A,k+1}^{E_1:\alpha_1}$.
Similarly one can argue that the edge $BA$ (one between levels $k+1$ and $k+2$) in $T_{A,k+1}^{E_1:\alpha_1}$ 
is same as the corresponding edge $BA$ in $T_{A,k+1}^{\alpha}$. 
Thus, $\bar{\cal A}$ can ensure that 
$T_{A,k+1}^{\alpha} \sim T_{A,k+1}^{E_{1}:\alpha_{1}}$. Since it is true for all values of $k$,
we have $T_{A}^{\alpha} \sim T_{A}^{E_{1}:\alpha_{1}}$. 

The proof for $V_{B}^{\alpha} \sim V_{B}^{E_1:\alpha_{1}}$ follows on very similar lines, 
we omit the details. \qed

\begin{lemma}\label{lbc3}
In execution $\alpha$, players $A$ and $B$ output $0$.
\end{lemma}
\emph{Proof}: From Lemma~\ref{lbc2}, it follows that player $A$ 
cannot distinguish execution $E_1$ of scenario $\alpha_{1}$ 
from execution $\alpha$ (dubbed as $E_1:\alpha_1 \stackrel{A}{\sim} \alpha$). 
Similarly, to player $B$ execution $E_1$ of scenario $\alpha_{1}$ 
is indistinguishable from $\alpha$ ($E_1:\alpha_1 \stackrel{B}{\sim} \alpha$). 
In $E_1:\alpha_1$, as per the definition of ABG [Definition~\ref{defABGCompose}] both 
$A$ and $B$ will decide on $0$. Since, $E_1:\alpha_1 \stackrel{A}{\sim} \alpha$ and 
$E_1:\alpha_1 \stackrel{B}{\sim} \alpha$, in $\alpha$ too $A$ and $B$ will decide on $0$. 
(We are able to make claims about the output of $A$ and $B$ in $\alpha$ as they 
cannot distinguish $E_1:\alpha_{1}$ from $\alpha$. Thus, by analysing their output in 
$E_1:\alpha_{1}$, we can determine their output in $\alpha$.) \qed \\

\noindent Adversary's strategy in $E_1:\alpha_{2}$ --\\
(Recall that adversary Byzantine corrupts $A$ only in $E_1:\alpha_2$)
\begin{enumerate}
\item \emph{Send outgoing messages of round i:} Based on the messages received 
during round $i-1$, $\bar{\cal{A}}$ decides on the messages to be sent in round 
$i$. For round $i$, $i\geq 1$, $\bar{\cal{A}}$ sends to $B$ what an honest $A$ 
would have sent to $B$ in execution $E_1:\alpha_1$. Formally, $\bar{\cal{A}}$ 
constructs $msg_{i}^{E_1:\alpha_{2}}(A,B)_{A}$ such that 
$msg_{i}^{E_1:\alpha_{2}}(A,B)_{A} = msg_{i}^{E_1:\alpha_{1}}(A,B)_{A}$.
Likewise, $\bar{\cal{A}}$ sends to $C$ what an honest $A$ would have sent to 
$C$ in execution $E_1:\alpha_3$. Formally, $\bar{\cal{A}}$ constructs 
$msg_{i}^{E_1:\alpha_{2}}(A,C)_{A}$ such that 
$msg_{i}^{E_1:\alpha_{2}}(A,C)_{A} = msg_{i}^{E_1:\alpha_{3}}(A,C)_{A}$.
Since $A$ is Byzantine faulty in $E_1:\alpha_{2}$, $\bar{\cal{A}}$
can always send the above stated messages.
\item \emph{Receive incoming messages of round i:} $\bar{\cal{A}}$ obtains messages $msg_{i}^{E_1:\alpha_{2}}(B,A)_{B}$ and $msg_{i}^{E_1:\alpha_{2}}(C,A)_{C}$ via $A$. 
\end{enumerate}

\begin{lemma}\label{lbc4}
$\bar{\cal A}$ can ensure $V_{B}^{\alpha} \sim V_{B}^{E_1:\alpha_{2}}$ and $V_{C}^{\alpha} \sim V_{C}^{E_1:\alpha_{2}}$.
\end{lemma}
\emph{Proof}: Using induction on $i$, we show that for any $i$, $T_{B,i}^{\alpha} \sim T_{B,i}^{E_1:\alpha_{2}}$. This 
implies $T_{B}^{\alpha} \sim T_{B}^{E_1:\alpha_{2}}$. From Equation~\ref{eqn5} it then then follows that $V_{B}^{\alpha} \sim V_{B}^{E_1:\alpha_{2}}$. \\

\noindent {\tt Base Case:}\\ 
\noindent \underline{$i=1$}\\
Consider $T_{B,1}^{\alpha}$ and $T_{B,1}^{E_1:\alpha_{2}}$ are shown in Figure~\ref{trees_0_2_1}. 
$C$ does not have any input in either $\alpha$ or $E_1:\alpha_{2}$. 
Thus, ${\cal I}_{C}^{\alpha} = {\cal I}_{C}^{E_1:\alpha_{2}}$ holds trivially. Further, 
$C$ starts with same secret key and executes same code in $\alpha$ and $E_1:\alpha_{2}$ i.e 
${\cal SK}_{C}^{\alpha} = {\cal SK}_{C}^{E_1:\alpha_{2}}$ and ${\cal \theta}_{C}^{\alpha} 
= {\cal \theta}_{C}^{E_1:\alpha_{2}}$. Thus, it will send same messages to $B$ in round 1 of 
$\alpha$ and $E_1:\alpha_{2}$ i.e. $msg_{1}^{E_1:\alpha_{2}}(C,B)_{C} = msg_{1}^{\alpha}(C,B)_{C}$. 
Since $A$ is Byzantine corrupt in $E_1:\alpha_{2}$, $\bar{\cal A}$ can ensure that 
$msg_{1}^{E_1:\alpha_{2}}(A,B)_{A} = msg_{1}^{\alpha}(A,B)_{A}$. Thus, 
$T_{B,1}^{\alpha} \sim T_{B,1}^{E_1:\alpha_{2}}$.

\begin{figure}[htbp]
\begin{center}
\input{subtree_0_2_1.pstex_t}
\caption{$T_{B,1}^{\alpha}$ and $T_{B,1}^{E_1:\alpha_{2}}$.}
\label{trees_0_2_1}
\end{center}
\end{figure}
\vspace{-10pt}

\noindent \underline{$i=2$}\\
We now argue that the similarity holds for round 2 as well. Consider $T_{B,2}^{\alpha}$ 
and $T_{B,2}^{E_1:\alpha_{2}}$ as shown Figure~\ref{trees_0_2_2}. 
Now, $B$ does not have any input in either $\alpha$ or $E_1:\alpha_{2}$, 
thus ${\cal I}_{B}^{\alpha} = {\cal I}_{B}^{E_1:\alpha_{2}}$ is trivially true.
Further, ${\cal SK}_{B}^{\alpha} = {\cal SK}_{B}^{E_1:\alpha_{2}}$ and 
${\cal \theta}_{B}^{\alpha} = {\cal \theta}_{B}^{E_1:\alpha_{2}}$. Hence, 
$msg_{1}^{E_1:\alpha_{2}}(B,A)_{B} = msg_{1}^{\alpha}(B,A)_{B}$ and 
$msg_{1}^{E_1:\alpha_{2}}(B,C)_{B} = msg_{1}^{\alpha}(B,C)_{B}$. Likewise,
$C$ does not have any input in $\alpha$ and $E_1:\alpha_{2}$, hence, 
${\cal I}_{C}^{\alpha} = {\cal I}_{C}^{E_1:\alpha_{2}}$. Also, 
${\cal SK}_{C}^{\alpha} = {\cal SK}_{C}^{E_1:\alpha_{2}}$ and 
${\cal \theta}_{C}^{\alpha} = {\cal \theta}_{C}^{E_1:\alpha_{2}}$. Hence, 
$msg_{1}^{E_1:\alpha_{2}}(C,A)_{C} = msg_{1}^{\alpha}(C,A)_{C}$. 
Since, $A$ is Byzantine corrupt in $E_1:\alpha_2$, $\bar{\cal A}$ can 
ensure that $msg_{1}^{\alpha}(A^\prime,C)_{A^\prime} = 
msg_{1}^{E_1:\alpha_{2}}(A,C)_{A}$. 

\begin{figure}[htbp]
\begin{center}
\input{subtree_0_2_2.pstex_t}
\caption{ $T_{B,2}^{\alpha}$ and $T_{A,2}^{E_1:\alpha_{1}}$.}
\label{trees_0_2_2}
\end{center}
\end{figure}

On similar lines, one can get $msg_{2}^{E_1:\alpha_{2}}(C,B)_{C} = msg_{2}^{\alpha}(C,B)_{C}$.
Since, $A$ is Byzantine corrupt in $E_1:\alpha_2$, $\bar{\cal A}$ can ensure that 
$msg_{2}^{E_1:\alpha_{2}}(A,B)_{A} = msg_{2}^{\alpha}(A,B)_{A}$. Thus, 
$T_{B,2}^{\alpha} \sim T_{B,2}^{E_1:\alpha_{2}}$. \\

\noindent {\tt Induction hypothesis:} Let it be true for all rounds upto $k$ i.e. $\forall i, ~i \leq k$, $T_{B,i}^{\alpha} \sim T_{B,i}^{E_{1}:\alpha_{2}}$. Likewise $\forall i, ~i \leq k$, $T_{C,i}^{\alpha} \sim T_{C,i}^{E_{1}:\alpha_{2}}$\\

\noindent {\tt Induction step:} We now show that the similarity holds for round $k+1$ too i.e. $T_{B,k+1}^{\alpha} \sim T_{B,k+1}^{E_{1}:\alpha_{2}}$. \\

\begin{figure}[htbp]
\begin{center}
\input{induction_sub_tree_2.pstex_t}
\caption{$T_{B,k+1}^{\alpha}$ and $T_{B,k+1}^{E_1:\alpha_{2}}$.}
\label{induction_sub_tree_2}
\end{center}
\end{figure}
\vspace{-10pt}

Consider $T_{B,k+1}^{\alpha}$ and $T_{B,k+1}^{E_1:\alpha_{2}}$ 
as shown in Figure~\ref{induction_sub_tree_2}. Consider the edges between level 
$k$ and $k+1$. From the induction hypothesis, we have $\forall j\leq k$, 
$T_{B,j}^{\alpha} \sim T_{B,j}^{E_1:\alpha_{2}}$. 
Now, ${\cal I}_{B}^{\alpha} = {\cal I}_{B}^{E_1:\alpha_{2}}$, 
${\cal SK}_{B}^{\alpha} = {\cal SK}_{B}^{E_1:\alpha_{2}}$, 
${\cal \theta}_{B}^{\alpha} = {\cal \theta}_{B}^{E_1:\alpha_{2}}$.
Thus, $B$ sends same messages to $A$ in round $k$ of both the executions i.e. 
$msg_{k}^{E_1:\alpha_{2}}(B,A)_{B} = msg_{k}^{\alpha}(B,A)_{B}$. 
Thus, edge $BA$ (between levels $k$ and $k+1$) is same in both the trees. 
Likewise, $B$ sends same messages to $C$ in round $k$ of both the executions 
i.e. $msg_{k}^{E_1:\alpha_{2}}(B,C)_{B} = msg_{k}^{\alpha}(B,C)_{B}$. 
Thus, edge $BC$ (between levels $k$ and $k+1$) is same in both the trees. 

From the induction hypothesis we have $\forall j\leq k$, $T_{C,j}^{\alpha} 
\sim T_{C,j}^{E_1:\alpha_{2}}$. Since, ${\cal I}_{C}^{\alpha} = {\cal I}_{C}^{E_1:\alpha_{2}}$, 
${\cal SK}_{C}^{\alpha} = {\cal SK}_{C}^{E_1:\alpha_{2}}$ and
${\cal \theta}_{C}^{\alpha} = {\cal \theta}_{C}^{E_1:\alpha_{2}}$, therefore,
$C$ sends same messages to $A$ in round $k$ of both the executions i.e. 
$msg_{k}^{E_1:\alpha_{2}}(C,A)_{C} = msg_{k}^{\alpha}(C,A)_{C}$. Hence, 
edge $CA$ (between levels $k$ and $k+1$) is same in both the trees.

Using same arguments as in preceding two paragraphs, we get 
$C$ sends same messages to $B$ in round $k+1$ of both the executions i.e. 
$msg_{k+1}^{E_1:\alpha_{2}}(C,B)_{C} = msg_{k+1}^{\alpha}(C,B)_{C}$. Hence, 
edge $CB$ (between levels $k+1$ and $k+2$) is same in both the trees. Now, given 
$A$ is Byzantine faulty in $E_1:\alpha_{2}$, $\bar{\cal A}$ sends that $msg_{k+1}^{E_1:\alpha_{2}}(A,B)_{A}$
which ensures $msg_{k+1}^{E_1:\alpha_{2}}(A,B)_{A} = msg_{k+1}^{\alpha}(A,B)_{A}$. 
Thus, the edge $AB$ (between levels $k$ and $k+1$) is same in both the trees. 

The proof for $V_{C}^{\alpha} \sim V_{C}^{E_1:\alpha_{2}}$ is nearly a repetition 
of the above arguments. Details omitted. \qed

\begin{lemma}\label{lbc5}
In execution $\alpha$, output of $B$ will be same output of $C$.
\end{lemma}
\emph{Proof}: From Lemma~\ref{lbc4}, we get that player $B$ 
cannot distinguish execution $E_1:\alpha_{2}$ 
from $\alpha$ (dubbed as $E_1:\alpha_2 \stackrel{B}{\sim} \alpha$). 
Similarly, to player $C$ execution $E_1:\alpha_{2}$ 
is indistinguishable from $\alpha$ ($E_1:\alpha_2 \stackrel{B}{\sim} \alpha$). 
Since, the General ($A$) is Byzantine corrupt in $E_1:\alpha_2$, from the definition 
of ABG [Definition~\ref{defABGCompose}], in $E_1:\alpha_2$, 
$B$ and $C$ must have the same output. Then, so must $B$ and $C$ in $\alpha$. \qed \\

\noindent Adversary's strategy in $E_1:\alpha_{3}$ --\\
(Recall that in $\alpha_3$ $\bar{\cal{A}}$ Byzantine corrupts $B$ in $E_1$ and $A$ in $E_2$.)
\begin{enumerate}
\item \emph{Send outgoing messages of round i:} Based on the messages received during round $i-1$, $\bar{\cal{A}}$ decides on the messages to be sent in round $i$. For round 1, $\bar{\cal{A}}$ sends to $C$ what an honest $B$ would have sent to $C$ in execution $E_1:\alpha_2$. For $i\geq 2$, $\bar{\cal{A}}$ authenticates $msg_{i-1}^{E_1:\alpha_{3}}(C,B)_{C}$ using $B$'s key and sends it to $A$. For $msg_{i-1}^{E_1:\alpha_{3}}(A,B)_{A}$, $\bar{\cal{A}}$ examines the message. If the message has not been authenticated by $C$ even once, it implies that the message has not yet been seen by $C$. Then, $\bar{\cal{A}}$ authenticates and sends a message to $C$ which is same as what $B$ would have sent to $C$ in round $i$ of execution $E_1:\alpha_{2}$. Formally, $\bar{\cal{A}}$ constructs $msg_{i-1}^{E_1:\alpha_{3}}(A,B)_{A}$ such that $msg_{i-1}^{E_1:\alpha_{3}}(A,B)_{A} \sim msg_{i-1}^{E_1:\alpha_{2}}(A,B)_{A}$. $\bar{\cal{A}}$ then authenticates $msg_{i-1}^{E_1:\alpha_{1}}(A,B)_{A}$ using $B$'s key and sends it to $C$. If the message has been authenticated by $B$ even once, $\bar{\cal{A}}$ simply authenticates $msg_{i-1}^{E_1:\alpha_{1}}(A,B)_{A}$ using $B$'s key and sends it to $C$.

\item \emph{Receive incoming messages of round i:} $\bar{\cal{A}}$ obtains messages $msg_{i}^{E_1:\alpha_{3}}(A,B)_{A}$ and $msg_{i}^{E_1:\alpha_{3}}(C,B)_{C}$ via $B$. Likewise, via $A$, $\bar{\cal{A}}$ obtains messages $msg_{i}^{E_1:\alpha_{3}}(B,A)_{B}$ and $msg_{i}^{E_1:\alpha_{3}}(C,A)_{C}$. 
\end{enumerate}

\begin{lemma}\label{lbc6}
$\bar{\cal A}$ can ensure $V_{C}^{\alpha} \sim V_{C}^{E_1:\alpha_{3}}$ and $V_{A^\prime}^{\alpha} \sim V_{A}^{E_1:\alpha_{3}}$.
\end{lemma}
\emph{Proof}: Owing to the symmetry of $\cal{L}$, the proof is very similar to the proof of Lemma~\ref{lbc2}. 
Details omitted. \qed

\begin{lemma}\label{lbc7}
In execution $\alpha$, players $C$ and $A^\prime$ output $1$.
\end{lemma}
\emph{Proof}: From Lemma~\ref{lbc6}, we have $E_1:\alpha_3 \stackrel{C}{\sim} \alpha$
and $E_1:\alpha_3 \stackrel{A^\prime}{\sim} \alpha$.
From Definition~\ref{defABGCompose}, 
$A$ and $C$ will output $1$ in $E_1:\alpha_3$.
Then, so must $A^\prime$ and $C$ in $\alpha$. \qed

\subsection{Finale}\label{MainCourse}
We now proceed to proving the main result of this work. 

\begin{theorem}[Main Theorem]\label{complete2}
 ABG over $n$ players, tolerating a $t$-adversary, can be 
 self-composed in parallel for any number of executions if and only if $n \geq 2t$.
\end{theorem}
\emph{Proof}: By combining Lemma~\ref{MainLemma1} and Lemma~\ref{MainLemma2}. \qed

\begin{lemma}\label{MainLemma1}
If $n < 2t$, then there does not exist any ABG protocol that self-composes in parallel even twice $(\Delta_{2})$
over a network of $n$ players, tolerating a $t$-adversary.
\end{lemma}
\emph{Proof}: Our proof demonstrates that if $n \leq 2t_1 +min(t_1,t_2)$, $t_2 > 0$, 
then there does not exist any $\Delta_{2}$, over a network 
(${\cal N^\prime}$) of $n$ players tolerating a ($t_1$,$t_2$)-adversary. Here, $t_1$ and $t_2$ are the number of 
players that the adversary can corrupt in the two parallel executions $E_1$ 
and $E_2$, respectively, of $\Delta_{2}$ such that $t_1$+$t_2 \leq t$ (dubbed as
($t_1$,$t_2$)-adversary). Substituting $t_{1} = t-1$ and $t_{2}=1$ in
$n \leq 2t_1 + min(t_1, t_2)$, gives the impossibility of $\Delta_{2}$ when $n < 2t$. 
Hence, the Lemma.

To show the impossibility of $\Delta_{2}$ when 
$n \leq 2t_1 +min(t_1,t_2)$, $t_2 > 0$, we assume otherwise and arrive 
at a contradiction. For the purpose of contradiction, we assume the existence 
of $\Delta_{2}$ over ${\cal N^\prime}$ such that $n \leq 2t_1 +min(t_1,t_2)$, $t_2 > 0$, 
tolerating a ($t_1$,$t_2$)-adversary. Using $\Delta_{2}$, 
we construct a protocol $\Pi_{2}$ over a network ${\cal N}$
of $3$ nodes, $\mathbb{P}$=\{$A$,$B$,$C$\}, that tolerates an adversary
basis $\bar{\cal A}=\{((C),(A)); ((A),(\emptyset)); ((B),(A))\}$. 
But this contradicts Theorem~\ref{5theorem_motiv}, hence 
the assumed protocol $\Delta_{2}$ cannot exist. 

Construction of $\Pi_{2}$ from $\Delta_{2}$ is as follows:
partition the $n$ players in $\cal N^\prime$ into three, mutually
disjoint, non-empty sets $I_{A}$, $I_{B}$ and $I_{C}$ such that
$|I_{A}| \leq min(t_1, t_2)$, $|I_{B}| \leq t_1$ and $|I_{C}| \leq
t_1$. Since $n \leq 2t_1 + min(t_1, t_2)$, such a partitioning is
always possible. The edges in $\cal N^\prime$ can then be considered as
bundle of edges between the sets $I_{A}$, $I_{B}$ and $I_{C}$. Let
$E_1$, $E_2$ be two parallel executions of $\Delta_{2}$. 
Since $\Delta_{2}$ tolerates $(t_1, t_2)$-adversary, then $\Delta_{2}$
will tolerate an adversary, ${\cal A}$, characterised 
by adversary basis $\{((I_{C}),(I_{A})); ((I_{A}),(\emptyset));
((I_{B}),(I_{A}))\}$. Let the corresponding parallel executions of
$\Pi_{2}$ be $E_1^\prime$ and $E_2^\prime$. Player $i$, $i \in \{A,B,C\}$,
in execution $E_l^\prime$, $l \in \{1,2\}$, simulates all the players 
in set $I_i$ in execution $E_l$. W.l.o.g, let the honest, passively corrupt and 
Byzantine faulty players in $E_l^\prime$ simulate the honest, 
passively corrupt and Byzantine faulty players respectively in $E_l$.

Player $i$ in $E_l^\prime$ simulates players in $I_i$ in $E_l$ 
as follows: player $i$ keeps a track of the states of all 
the players in $I_i$. It assigns its input value to every 
member of $I_i$ and emulates the steps of all the players 
in $I_i$ as well as the messages communicated between every 
pair of players in $I_i$. If any player in $I_i$ sends a message to any player 
in $I_{j}$, $j \in \{A, B, C\}, i \neq j$, then player $i$ sends exactly the 
same message to player $j$. If any player in $I_{i}$ terminates then so does 
player $i$. If any player in $I_{i}$ decides on a value $v$, then so does 
player $i$.

We, now, show that if $\Delta_{2}$ satisfies Definition~\ref{defABGCompose}
tolerating ${\cal A}$, then so does $\Pi_{2}$ tolerating 
$\bar{\cal A}=\{((C),(A)); ((A),(\emptyset)); ((B),(A))\}$.
Let $i$ and $j$, ($i \neq j$), be two non-faulty players (honest or 
passively corrupt but not Byzantine faulty) in execution 
$E_l^\prime$ of $\Pi_{2}$. Player $i$ (likewise $j$) 
simulates at least one player in $I_i$ ($I_j$) in execution $E_l$. 
Since both $i,j$ are non-faulty in $E_l^\prime$, then so 
are all the players in $I_i, I_j$ in $E_l$. If the \emph{General} 
is non-faulty in $E_l^\prime$ and starts with a value $v$, then in 
$E_l$ too, the General is non-faulty and starts with a value $v$. 
Hence, as per the definition of ABG [Definition~\ref{defABGCompose}], 
all the players in $I_i$, $I_j$ in execution $E_l$ must decide on 
value $v$. Then, so should players $i,j$ in $E_l^\prime$. If the 
\emph{General} is faulty in $E_l^\prime$, then so is the \emph{General} in $E_l$. 
As per the definition of ABG all the players in $I_i$, $I_j$ in 
execution $E_l$ must have the same output. Then, so should
players $i,j$ in $E_l^\prime$. This implies $\Pi_{2}$ satisfies 
Definition~\ref{defABGCompose} tolerating $\bar{\cal A}$, contradicting Theorem~\ref{5theorem_motiv}. \qed\\

We now show that the bound of $n < 2t$ is tight. For this we present 
a protocol -- $EIGPrune^+$ (Figure~\ref{EIG_protocol+}) and prove that 
if $n \geq 2t$, then $EIGPrune^+$ remains a valid ABG protocol under
any number parallel self-compositions (Lemma~\ref{MainLemma2}). $EIGPrune^+$ 
is based on a sequence of transformations on EIG tree~\cite{BarDoDwStr87}.
\cite[page 108]{NancyBook} gives an excellent discussion on the construction 
of EIG tree. $EIGPrune^+$ is essentially same as {\em EIGPrune} 
protocol~\cite{GGBS10}\footnote{For the benefit of the reader we reproduce 
{\em EIGPrune}, as proposed by Gupta {\em et al.}, in Appendix~\ref{EIGp}.}, 
barring two minor additions -- (i) Each concurrent execution of the protocol 
is augmented with a {Unique Session Identifier}(USID). (ii) Non-faulty players 
in any concurrent execution reject any message that does not carry a valid 
USID. We remark that like $EIGPrune$, $EIGPrune^+$ is also exponential 
in the number of messages. However, owing to the simplicity and intuitive 
appeal of the protocol  we present the same as it makes the discussion very 
lucid. The exponential nature of the protocol is not a serious concern
as using well known techniques in literature~\cite{BarDoDwStr87}, it can be 
converted into an efficient protocol. \\

\begin{figure}[h]
\begin{center}
\begin{tabular}{|c|}
\hline\\
\parbox{6.5in}
{
\vspace{-5pt}
\begin{small}
In every concurrent execution:
\begin{enumerate}
\item The General ${\cal G}$ send his value to every player. 
\item On receiving this value from ${\cal G}$, every player 
assumes it be his input  value and exchanges messages with others as per {\em EIGStop} 
protocol~\cite[page 103]{NancyBook} for $t+1$ rounds. 
\item At the end of $t+1$ rounds of {\em EIGStop} protocol, 
player $p_i$ discards any messages that does not have a 
valid authentication or USID and 
invokes {\bf Prune(EIG)} [Definition~\ref{4prune}]. 
\item Player $p_i$ applies the following decision rule -- take
majority of the values at the first level (i.e. all the
nodes with labels $l$ such that $l \in {\mathbb P}$) of 
its {\em EIG} tree. If a majority exists, player $p_i$ 
decides on that value; else, $p_i$ decides on {\em default value}, $v_0$.
\end{enumerate}
\end{small}
\vspace{5pt}
} \\ 
\hline
\end{tabular}
\end{center}
\vspace{-15pt}
\caption{\label{EIG_protocol+}{$EIGPrune^+$ Protocol}}
\end{figure}
\vspace{-10pt}

\begin{definition}[{\bf Prune}(EIG)]\label{4prune}
It takes an EIG tree as an input and deletes subtrees
say ${s_j}^i$ (subtree rooted at a node whose's label is
$j$ in EIG tree of player $i$) as given in the sequel. For each subtree
${s_j}^i$, where label $j \in {\mathbb P}$, a set $W_{j}$ is
constructed that contains all distinct values that ever appears in
${s_j}^i$. If $|W_{j}| > 1$, ${s_j}^i$ is deleted and
modified EIG tree is returned.
\end{definition}

Our proof of correctness of $EIGPrune^+$ is based on the idea developed by 
Lindell \emph{et al.}\/~\cite{LLR02:OtCoABA}, wherein, the security of a 
protocol in the concurrent setting is reduced to the security 
of the protocol in the stand alone setting. In our case, to ensure that 
this reduction is correct, we must account for the possibility of the
players being (implicitly) passively corrupt in an execution (from the observation made in section~\ref{subsec22}).
For this reason, we use a variant of ABG - christened as $ABG_{mix}$, 
proposed by Gupta \emph{et al.}~\cite{GGBS10}, as the stand alone setting for our reduction.

Gupta {\em et al.} studied (stand alone) ABG in the presence 
of a mixed adversary that can corrupt up to any $t_b$ players 
actively and up to another $t_p$ players passively 
(dubbed as ($t_b$,$t_p$)-adversary). The adversary, thus, can 
forge the signatures of all $t_b+t_p$ players. Further, Gupta {\em et al.}
require all the passively corrupt players to always output
a value same as the output of the honest players. They prove that (stand alone) 
$ABG_{mix}$ over a completely connected synchronous network of 
$n$ players tolerating a ($t_b$,$t_p$)-adversary, $t_p >0$, is 
solvable if and only if $n>2t_b+min(t_b,t_p)$. It is easy to see that 
like $EIGPrune$, $EIGPrune^+$ is also a correct protocol for $ABG_{mix}$.
By substituting $t_b=t-1$ and $t_p=1$, their result can be extended to 
achieve a bound of $n \geq 2t$. 

\begin{lemma}\label{MainLemma2}
$EIGPrune^+$ over $n$ players, tolerating a $t$-adversary, is a valid ABG 
protocol that self-composes for any number of parallel executions if $n \geq 2t$.
\end{lemma}
\emph{Proof}: Let $\Psi(id_1), \Psi(id_2), \ldots ,\Psi(id_l)$\footnote{$\Psi(id_i)$ denotes 
the use of USID in the $i^{th}$ concurrent execution of $\Psi$.} be $l$ parallel 
executions of $EIGPrune^{+}$. For the purpose of contradiction assume 
that there exists an adversary, ${\cal A}$, that attacks these $l$ 
parallel executions and succeeds in execution $\Psi(id_i)$, for some $i \in (1,l)$. Using 
${\cal A}$ we construct an adversary ${\cal A^\prime}$ that is 
bound to succeed in the (stand alone) execution, $\varphi$, of {\em EIGPrune} for $ABG_{mix}$. This 
contradicts the results of Gupta \emph{et al.}, hence the Lemma.

Our construction of ${\cal A^\prime}$ requires ${\cal A^\prime}$ 
to internally simulate the parallel executions 
$\Psi(id_1), \Psi(id_2), \ldots,$\\$\Psi(id_l)$. 
For this to happen, ${\cal A^\prime}$ must be able to simulate 
the signatures generated by the honest players in these executions. 
To facilitate the same we assume ${\cal A^\prime}$ is given access 
to all the oracles\footnote{Players use the signature scheme,
(($Gen$,$S_{id}$,$V_{id}$),$S_{\neg id}$), developed by Lindell
\emph{et al.}\/~\cite{LLR02:OtCoABA}. We present a brief overview 
of the same in Appendix~\ref{SigSch}.}
$S_{\neg id}(sk_{1},\cdot), \ldots, S_{\neg id}(sk_{n},\cdot)$. 
Further, to ensure that the above simulation is perfect we 
augment the stand alone execution, $\varphi$, of {\em EIGPrune} 
for $ABG_{mix}$ with USID -- $\varphi(id)$. It is easy to see 
that this augmentation has no bearing on the correctness 
of {\em EIGPrune}. Formally, for $ABG_{mix}$, if $\varphi$ is a 
correct execution of {\em EIGPrune}, then so is $\varphi(id)$.

Construction of $\cal{A}^\prime$ is as follows: 
$\cal{A}^\prime$ internally incorporates $\bar{\cal A}$ and attacks
$\varphi(id)$ as given in the sequel. $\cal{A}^\prime$ selects an execution $i$,
$i \in (1,l)$, and sets $id$=$id_i$. Then, $\cal{A}^\prime$
invokes $\bar{\cal A}$ and simulates the concurrent executions of 
$\Psi(id_1) \ldots \Psi(id_l)$ for $\bar{\cal A}$. $\cal{A}^\prime$  
does this by playing the roles of the non-faulty players in all
the executions but $\Psi(id)$. Since $\cal{A}^\prime$  has 
access to the signing oracles $S_{\neg id}(sk_1,\cdot), \ldots, S_{\neg id}(sk_n,\cdot)$, 
it can generate signature on behalf of honest players in all the
executions $\Psi(id_j)$, $j \neq i$. In $\Psi(id)$, $\cal{A}^\prime$  
externally interacts with the non-faulty players and passes messages 
between them and $\bar{\cal A}$. $\cal{A}^\prime$ interacts with the players
in $\Psi(id_i)$ in exactly the same manner as $\bar{\cal A}$ interacts with 
players in $\Psi(id_i)$. Note that this is possible because if 
$\bar{\cal A}$ forges messages on behalf of some player in $\Psi(id_i)$
by active corrupting this player in $\Psi(id_j)$, $j \neq i$; then 
$\cal{A}^\prime$ can do the same by passively corrupting this particular 
player in $\Psi(id)$. Since, $\cal{A}^\prime$ never queries the oracle for messages whose prefix is $id$.
Therefore, the emulation by $\cal{A}^\prime$of the concurrent executions for $\bar{\cal A}$ 
is perfect. Thus, if $\bar{\cal A}$ succeeds in breaking $\Psi(id)$, 
then $\cal{A}^\prime$ will break $\varphi(id)$. \qed 

\section{Closing Remarks}\label{Conclusion}

In this paper we argue for the need of a better model for studying
self composition of ABG protocols. We propose a new model to study
composition of ABG protocols and show that, in this model, unique session
identifiers aid in improving the fault-tolerance of ABG protocols
(that compose in parallel) but from $n>3t$ only to $n \geq 2t$. Note that, in the stand
alone setting, ABG is possible for $n>t$. Thus surprisingly, USID's may
not always achieve their goal of truly {\em separating}\/ the
protocol's execution from its environment to the fullest extent.
 However, for most functionalities, USID's indeed achieve their objective, 
as is obvious from Universal Composability (UC) theorem~\cite{C01:UCS:aNPfCP}.

Besides proving (im-)possibility results for self composition of ABG, our
work also brings to forefront a few minor, yet interesting and undesirable
properties of UC framework. It will nice to see if one can fine tune UC
framework to this end. Further, with respect to composition of ABG protocols, 
we show that the worst-case 
adversary (with respect to a given execution) is not the one that corrupts 
players at full-throttle across all protocols running concurrently in the 
network. There may be several other problems apart from ABG wherein
similar anomaly holds. It is an intriguing open question to 
characterize the set of all such problems. Further, from our results
of $n \geq 2t$,  it appears that studying self composition of ABG protocols over
general networks will be interesting in its own right.

\bibliography{composabilityTR}
\newpage
\appendix
\section{{{\em EIGPrune} Protocol}~\cite{GGBS10}}\label{EIGp}
\begin{figure}[h]
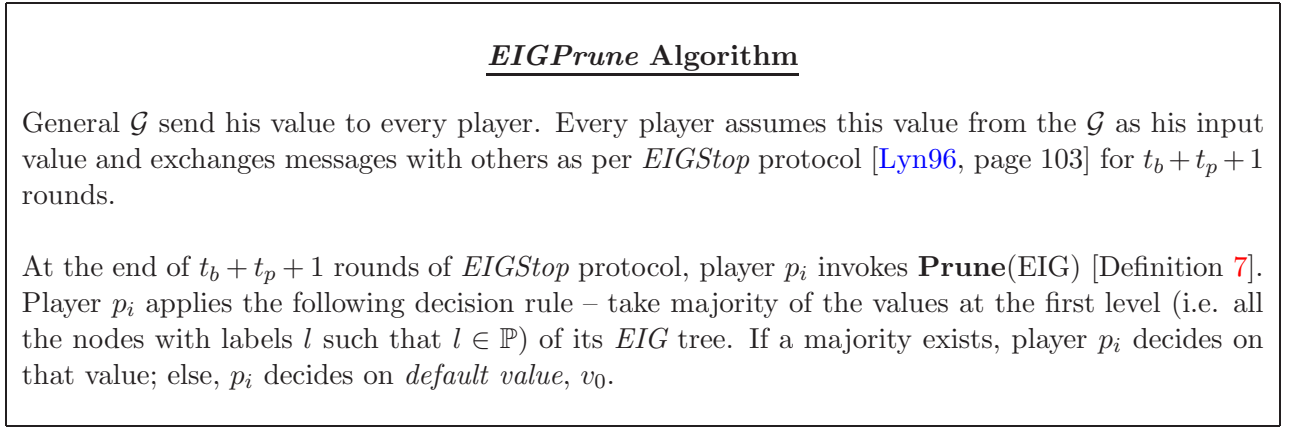

\begin{center}
\begin{tabular}{|c|}
\hline\\
\textbf{\underline{{\em EIGPrune} Algorithm}}\\ \\
\parbox{6.5in}{General ${\cal G}$ send his value to every player. 
Every player assumes this value from the ${\cal G}$ as his input
 value and exchanges messages with others as per {\em EIGStop}
 protocol~\cite[page 103]{NancyBook} for $t_b+t_p+1$ rounds. \\

At the end of $t_b+t_p+1$ rounds of {\em EIGStop} protocol, 
player $p_i$ invokes {\bf Prune}(EIG) [Definition~\ref{pruneDef}]. 
Player $p_i$ applies the following decision rule -- take
majority of the values at the first level (i.e. all the
nodes with labels $l$ such that $l \in {\mathbb P}$) of 
its {\em EIG} tree. If a majority exists, player $p_i$ 
decides on that value; else, $p_i$ decides on {\em default value}, $v_0$.
} \\ \\
\hline
\end{tabular}
\end{center}
\vspace{-15pt}
\caption{\label{4EIGPruneAlgo}{\em EIGPrune} algorithm}
\end{figure}

For the benefit of the reader, we now
reproduce the protocol proposed by Gupta {\emph et al.} 
(Figure~\ref{4EIGPruneAlgo}). It is obtained by a sequence of 
transformations on EIG tree~\cite{BarDoDwStr87}.  
\cite[page 108]{NancyBook} gives an excellent discussion on 
the construction of EIG tree. All the messages exchanged during 
the protocol are signed/authenticated. We only give the protocol
here. The proof of its correctness in stand alone setting for the 
problem of $ABG_{mix}$ can be found in~\cite{GGBS10} 

\vspace{-5pt}

\begin{definition}[{\bf Prune}(EIG)]\label{pruneDef}
  This method that takes an EIG tree as an input and deletes subtrees
  say ${subtree_j}^i$ (${subtree_j}^i$ refers to a subtree in $i$'s
  EIG tree such that the subtree is rooted at node whose's label is
  $j$) of $i's$ EIG tree as given in the sequel. For each subtree
  ${subtree_j}^i$, where label $j \in {\mathbb P}$, a set $W_{j}$ is
  constructed which contains all distinct values that ever appears in
  ${subtree_j}^i$. If $|W_{j}| > 1$, ${subtree_j}^i$ is deleted and
  modified EIG tree is returned.
\end{definition}

%
%
%

\section{Lindell \emph{et al.}'s Signature Scheme}\label{SigSch}
We present an overview of the signature scheme (($Gen$,$S_{id}$,$V_{id}$),$S_{\neg id}$)
developed by Lindell \emph{et al.}\/~\cite{LLR02:OtCoABA}. They define a signature scheme as ($Gen$,$S$,$V$) where
$S$,$V$ are are algorithms for signing and verification of any
message. $Gen$ is used to generate signature and verification keys for
a particular player (say $P_k$) and defined as a function: $(1)^{n}
\rightarrow$ ($v{k},s{k}$). A signature scheme is said to be a valid
one if honestly generated signatures are almost always
accepted. Formally, with non negligible probability, for every message
$m$, V($vk$,$m$,S($sk$,$m$)) = 1, where ($v{k},s{k}$) $\leftarrow$
$(1)^{n}$.  They model the valid signatures that adversary $\bar{\cal A}$
can obtain in a real attack via a signing oracle
$S(sk,\cdot)$. $\bar{\cal A}$ is defined to succeed in generating a forged
message $m^{*}$ if $\bar{\cal A}$ given $vk$, access to oracle
$S(sk,\cdot)$ can generate a pair $(m^{*}, \sigma^{*})$ such that if
$Q_{m}$ is the set of oracle queries made by $\bar{\cal A}$ then
V($vk$,$m^{*}$,$\sigma^{*}$) = 1 holds true if $m^{*} \not \in
Q_{m}$. A signature scheme is said to be existentially secure against
chosen-message attack if $\bar{\cal A}$ cannot succeed in forging a
signature with greater than non-negligible probability.  They further
model any information gained by $\bar{\cal A}$ from any query with another
oracle Aux($sk$,.). However, this oracle cannot generate any valid
signature but provides any other auxiliary information about the
query. They assume some scheme say ($Gen$,$S$,$V$) to be secure
against chosen-message attack and show how to construct a secure
scheme ($Gen$,$S_{id}$,$V_{id}$) from it where $S_{id}$($sk$,$m$) =
$S$($sk$,$id \circ m$) and $V_{id}$($vk$,$m$,$\sigma$) = $V$($vk$,$id
\circ m$,$\sigma$). For the new scheme they define the oracle
Aux($sk$,$\cdot$) = $S_{\neg id}$($sk$,$m$) where $S_{\neg
  id}$($sk$,$m$) = $S$($sk$,$m$) if the prefix of $m$ is not $id$ else
$S_{\neg id}$($sk$,$m$) = $\perp$. Further, they assume $\pi$ to be a
secure protocol for ABG using signature scheme ($Gen$,$S$,$V$). They
define modified protocol $\pi(id)$ to be exactly same as $\pi$ except
that it uses signature scheme ($Gen$,$S_{id}$,$V_{id}$) as defined
above. They further prove as to why (($Gen$,$S_{id}$,$V_{id}$),
$S_{\neg id}$) is secure against chosen-message attack. Intuition
behind the proof is the fact that if the prefix of $m \neq id$, then
$S_{\neg id}$($sk$,$m$) = $S$($sk$,$m$) which is of no help to the
adversary as any successful forgery must be prefixed with $id$ and all
oracle queries to $S_{\neg id}$ must be prefixed with $id^\prime \neq
id$. 

\end{document}

%% file: 1.pstex_t
\begin{picture}(0,0)%
\includegraphics{1.pstex}%
\end{picture}%
\setlength{\unitlength}{3947sp}%
\begingroup\makeatletter\ifx\SetFigFont\undefined%
\gdef\SetFigFont#1#2#3#4#5{%
  \reset@font\fontsize{#1}{#2pt}%
  \fontfamily{#3}\fontseries{#4}\fontshape{#5}%
  \selectfont}%
\fi\endgroup%
\begin{picture}(1528,1688)(70,-1169)
\put(904,372){\makebox(0,0)[lb]{\smash{{\SetFigFont{12}{14.4}{\rmdefault}{\mddefault}{\updefault}{\color[rgb]{0,0,0}$C$}%
}}}}
\put(1537,-1105){\makebox(0,0)[lb]{\smash{{\SetFigFont{12}{14.4}{\rmdefault}{\mddefault}{\updefault}{\color[rgb]{0,0,0}$B$}%
}}}}
\put( 85,-1105){\makebox(0,0)[lb]{\smash{{\SetFigFont{12}{14.4}{\rmdefault}{\mddefault}{\updefault}{\color[rgb]{0,0,0}$A$}%
}}}}
\end{picture}%

%% file: 2.pstex_t
\begin{picture}(0,0)%
\includegraphics{2.pstex}%
\end{picture}%
\setlength{\unitlength}{3947sp}%
\begingroup\makeatletter\ifx\SetFigFont\undefined%
\gdef\SetFigFont#1#2#3#4#5{%
  \reset@font\fontsize{#1}{#2pt}%
  \fontfamily{#3}\fontseries{#4}\fontshape{#5}%
  \selectfont}%
\fi\endgroup%
\begin{picture}(6907,3287)(251,-2768)
\put(1293,-31){\makebox(0,0)[lb]{\smash{{\SetFigFont{10}{12.0}{\rmdefault}{\mddefault}{\updefault}{\color[rgb]{0,0,0}$\Pi^\prime$}%
}}}}
\put(961,-620){\makebox(0,0)[lb]{\smash{{\SetFigFont{10}{12.0}{\rmdefault}{\mddefault}{\updefault}{\color[rgb]{0,0,0}$\Pi^\prime$}%
}}}}
\put(1897,-183){\makebox(0,0)[lb]{\smash{{\SetFigFont{10}{12.0}{\rmdefault}{\mddefault}{\updefault}{\color[rgb]{0,0,0}$A^\prime$}%
}}}}
\put(1029,-245){\makebox(0,0)[lb]{\smash{{\SetFigFont{10}{12.0}{\rmdefault}{\mddefault}{\updefault}{\color[rgb]{0,0,0}$C^\prime$}%
}}}}
\put(746,372){\makebox(0,0)[lb]{\smash{{\SetFigFont{10}{12.0}{\rmdefault}{\mddefault}{\updefault}{\color[rgb]{0,0,0}$B^\prime$}%
}}}}
\put(1144,-386){\makebox(0,0)[lb]{\smash{{\SetFigFont{10}{12.0}{\rmdefault}{\mddefault}{\updefault}{\color[rgb]{0,0,0}$C$}%
}}}}
\put(3721,-480){\makebox(0,0)[lb]{\smash{{\SetFigFont{10}{12.0}{\rmdefault}{\mddefault}{\updefault}{\color[rgb]{0,0,0}$C$}%
}}}}
\put(3247,-245){\makebox(0,0)[lb]{\smash{{\SetFigFont{10}{12.0}{\rmdefault}{\mddefault}{\updefault}{\color[rgb]{0,0,0}$C^\prime$}%
}}}}
\put(3106,372){\makebox(0,0)[lb]{\smash{{\SetFigFont{10}{12.0}{\rmdefault}{\mddefault}{\updefault}{\color[rgb]{0,0,0}$B^\prime$}%
}}}}
\put(5705,181){\makebox(0,0)[lb]{\smash{{\SetFigFont{10}{12.0}{\rmdefault}{\mddefault}{\updefault}{\color[rgb]{0,0,0}$C^\prime$}%
}}}}
\put(6556,181){\makebox(0,0)[lb]{\smash{{\SetFigFont{10}{12.0}{\rmdefault}{\mddefault}{\updefault}{\color[rgb]{0,0,0}$B^\prime$}%
}}}}
\put(287,-480){\makebox(0,0)[lb]{\smash{{\SetFigFont{10}{12.0}{\rmdefault}{\mddefault}{\updefault}{\color[rgb]{0,0,0}$A$}%
}}}}
\put(2655,-527){\makebox(0,0)[lb]{\smash{{\SetFigFont{10}{12.0}{\rmdefault}{\mddefault}{\updefault}{\color[rgb]{0,0,0}$A$}%
}}}}
\put(4317,-245){\makebox(0,0)[lb]{\smash{{\SetFigFont{10}{12.0}{\rmdefault}{\mddefault}{\updefault}{\color[rgb]{0,0,0}$A^\prime$}%
}}}}
\put(5366,-320){\makebox(0,0)[lb]{\smash{{\SetFigFont{10}{12.0}{\rmdefault}{\mddefault}{\updefault}{\color[rgb]{0,0,0}$A$}%
}}}}
\put(7143,-289){\makebox(0,0)[lb]{\smash{{\SetFigFont{10}{12.0}{\rmdefault}{\mddefault}{\updefault}{\color[rgb]{0,0,0}$A^\prime$}%
}}}}
\put(1379,-1072){\makebox(0,0)[lb]{\smash{{\SetFigFont{10}{12.0}{\rmdefault}{\mddefault}{\updefault}{\color[rgb]{0,0,0}$B$}%
}}}}
\put(3730,-1164){\makebox(0,0)[lb]{\smash{{\SetFigFont{10}{12.0}{\rmdefault}{\mddefault}{\updefault}{\color[rgb]{0,0,0}$B$}%
}}}}
\put(698,-1646){\makebox(0,0)[lb]{\smash{{\SetFigFont{10}{12.0}{\rmdefault}{\mddefault}{\updefault}{\color[rgb]{0,0,0}$C^\prime$}%
}}}}
\put(1550,-1646){\makebox(0,0)[lb]{\smash{{\SetFigFont{10}{12.0}{\rmdefault}{\mddefault}{\updefault}{\color[rgb]{0,0,0}$B^\prime$}%
}}}}
\put(3155,-1646){\makebox(0,0)[lb]{\smash{{\SetFigFont{10}{12.0}{\rmdefault}{\mddefault}{\updefault}{\color[rgb]{0,0,0}$C^\prime$}%
}}}}
\put(4051,-1646){\makebox(0,0)[lb]{\smash{{\SetFigFont{10}{12.0}{\rmdefault}{\mddefault}{\updefault}{\color[rgb]{0,0,0}$B^\prime$}%
}}}}
\put(5753,-1646){\makebox(0,0)[lb]{\smash{{\SetFigFont{10}{12.0}{\rmdefault}{\mddefault}{\updefault}{\color[rgb]{0,0,0}$C^\prime$}%
}}}}
\put(6649,-1646){\makebox(0,0)[lb]{\smash{{\SetFigFont{10}{12.0}{\rmdefault}{\mddefault}{\updefault}{\color[rgb]{0,0,0}$B^\prime$}%
}}}}
\put(2050,-2062){\makebox(0,0)[lb]{\smash{{\SetFigFont{10}{12.0}{\rmdefault}{\mddefault}{\updefault}{\color[rgb]{0,0,0}$A^\prime$}%
}}}}
\put(266,-2085){\makebox(0,0)[lb]{\smash{{\SetFigFont{10}{12.0}{\rmdefault}{\mddefault}{\updefault}{\color[rgb]{0,0,0}$A$}%
}}}}
\put(2776,-2073){\makebox(0,0)[lb]{\smash{{\SetFigFont{10}{12.0}{\rmdefault}{\mddefault}{\updefault}{\color[rgb]{0,0,0}$A$}%
}}}}
\put(4523,-2085){\makebox(0,0)[lb]{\smash{{\SetFigFont{10}{12.0}{\rmdefault}{\mddefault}{\updefault}{\color[rgb]{0,0,0}$A^\prime$}%
}}}}
\put(5374,-2085){\makebox(0,0)[lb]{\smash{{\SetFigFont{10}{12.0}{\rmdefault}{\mddefault}{\updefault}{\color[rgb]{0,0,0}$A$}%
}}}}
\put(7123,-2085){\makebox(0,0)[lb]{\smash{{\SetFigFont{10}{12.0}{\rmdefault}{\mddefault}{\updefault}{\color[rgb]{0,0,0}$A^\prime$}%
}}}}
\put(1594,-2713){\makebox(0,0)[lb]{\smash{{\SetFigFont{10}{12.0}{\rmdefault}{\mddefault}{\updefault}{\color[rgb]{0,0,0}$C$}%
}}}}
\put(605,-2713){\makebox(0,0)[lb]{\smash{{\SetFigFont{10}{12.0}{\rmdefault}{\mddefault}{\updefault}{\color[rgb]{0,0,0}$B$}%
}}}}
\put(3106,-2713){\makebox(0,0)[lb]{\smash{{\SetFigFont{10}{12.0}{\rmdefault}{\mddefault}{\updefault}{\color[rgb]{0,0,0}$B$}%
}}}}
\put(4100,-2713){\makebox(0,0)[lb]{\smash{{\SetFigFont{10}{12.0}{\rmdefault}{\mddefault}{\updefault}{\color[rgb]{0,0,0}$C$}%
}}}}
\put(5690,-2713){\makebox(0,0)[lb]{\smash{{\SetFigFont{10}{12.0}{\rmdefault}{\mddefault}{\updefault}{\color[rgb]{0,0,0}$B$}%
}}}}
\put(6700,-2713){\makebox(0,0)[lb]{\smash{{\SetFigFont{10}{12.0}{\rmdefault}{\mddefault}{\updefault}{\color[rgb]{0,0,0}$C$}%
}}}}
\put(6708,-871){\makebox(0,0)[lb]{\smash{{\SetFigFont{10}{12.0}{\rmdefault}{\mddefault}{\updefault}{\color[rgb]{0,0,0}$C$}%
}}}}
\put(5690,-871){\makebox(0,0)[lb]{\smash{{\SetFigFont{10}{12.0}{\rmdefault}{\mddefault}{\updefault}{\color[rgb]{0,0,0}$B$}%
}}}}
\end{picture}%

%% file: 3.pstex_t
\begin{picture}(0,0)%
\includegraphics{3.pstex}%
\end{picture}%
\setlength{\unitlength}{3947sp}%
\begingroup\makeatletter\ifx\SetFigFont\undefined%
\gdef\SetFigFont#1#2#3#4#5{%
  \reset@font\fontsize{#1}{#2pt}%
  \fontfamily{#3}\fontseries{#4}\fontshape{#5}%
  \selectfont}%
\fi\endgroup%
\begin{picture}(3801,2220)(1632,-1423)
\put(4842,-114){\makebox(0,0)[lb]{\smash{{\SetFigFont{14}{16.8}{\rmdefault}{\mddefault}{\updefault}{\color[rgb]{0,0,0}$A^\prime$}%
}}}}
\put(4153,578){\makebox(0,0)[lb]{\smash{{\SetFigFont{14}{16.8}{\rmdefault}{\mddefault}{\updefault}{\color[rgb]{0,0,0}$B^\prime$}%
}}}}
\put(2605,602){\makebox(0,0)[lb]{\smash{{\SetFigFont{14}{16.8}{\rmdefault}{\mddefault}{\updefault}{\color[rgb]{0,0,0}$C^\prime$}%
}}}}
\put(2026,-136){\makebox(0,0)[lb]{\smash{{\SetFigFont{14}{16.8}{\rmdefault}{\mddefault}{\updefault}{\color[rgb]{0,0,0}$A$}%
}}}}
\put(4195,-1006){\makebox(0,0)[lb]{\smash{{\SetFigFont{14}{16.8}{\rmdefault}{\mddefault}{\updefault}{\color[rgb]{0,0,0}$C$}%
}}}}
\put(2581,-1006){\makebox(0,0)[lb]{\smash{{\SetFigFont{14}{16.8}{\rmdefault}{\mddefault}{\updefault}{\color[rgb]{0,0,0}$B$}%
}}}}
\end{picture}%

%% file: induction_sub_trees_11.pstex_t
\begin{picture}(0,0)%
\includegraphics{induction_sub_trees_11.pstex}%
\end{picture}%
\setlength{\unitlength}{3947sp}%
\begingroup\makeatletter\ifx\SetFigFont\undefined%
\gdef\SetFigFont#1#2#3#4#5{%
  \reset@font\fontsize{#1}{#2pt}%
  \fontfamily{#3}\fontseries{#4}\fontshape{#5}%
  \selectfont}%
\fi\endgroup%
\begin{picture}(5373,2093)(-614,-1482)
\put(3508,-337){\makebox(0,0)[lb]{\smash{{\SetFigFont{12}{14.4}{\rmdefault}{\mddefault}{\updefault}{\color[rgb]{0,0,0}$A$}%
}}}}
\put(3862,-337){\makebox(0,0)[lb]{\smash{{\SetFigFont{12}{14.4}{\rmdefault}{\mddefault}{\updefault}{\color[rgb]{0,0,0}$C$}%
}}}}
\put(4256, 68){\makebox(0,0)[lb]{\smash{{\SetFigFont{12}{14.4}{\rmdefault}{\mddefault}{\updefault}{\color[rgb]{0,0,0}$C$}%
}}}}
\put(3775, 68){\makebox(0,0)[lb]{\smash{{\SetFigFont{12}{14.4}{\rmdefault}{\mddefault}{\updefault}{\color[rgb]{0,0,0}$B$}%
}}}}
\put(4231,-337){\makebox(0,0)[lb]{\smash{{\SetFigFont{12}{14.4}{\rmdefault}{\mddefault}{\updefault}{\color[rgb]{0,0,0}$B$}%
}}}}
\put(4039,464){\makebox(0,0)[lb]{\smash{{\SetFigFont{12}{14.4}{\rmdefault}{\mddefault}{\updefault}{\color[rgb]{0,0,0}$A$}%
}}}}
\put(4138,-691){\makebox(0,0)[lb]{\smash{{\SetFigFont{12}{14.4}{\rmdefault}{\mddefault}{\updefault}{\color[rgb]{0,0,0}$A$}%
}}}}
\put(4315,-691){\makebox(0,0)[lb]{\smash{{\SetFigFont{12}{14.4}{\rmdefault}{\mddefault}{\updefault}{\color[rgb]{0,0,0}$C$}%
}}}}
\put(4532,-691){\makebox(0,0)[lb]{\smash{{\SetFigFont{12}{14.4}{\rmdefault}{\mddefault}{\updefault}{\color[rgb]{0,0,0}$B$}%
}}}}
\put(4723,-691){\makebox(0,0)[lb]{\smash{{\SetFigFont{12}{14.4}{\rmdefault}{\mddefault}{\updefault}{\color[rgb]{0,0,0}$C$}%
}}}}
\put(4471,-337){\makebox(0,0)[lb]{\smash{{\SetFigFont{12}{14.4}{\rmdefault}{\mddefault}{\updefault}{\color[rgb]{0,0,0}$A$}%
}}}}
\put(3744,-691){\makebox(0,0)[lb]{\smash{{\SetFigFont{12}{14.4}{\rmdefault}{\mddefault}{\updefault}{\color[rgb]{0,0,0}$B$}%
}}}}
\put(3546,-691){\makebox(0,0)[lb]{\smash{{\SetFigFont{12}{14.4}{\rmdefault}{\mddefault}{\updefault}{\color[rgb]{0,0,0}$C$}%
}}}}
\put(3350,-691){\makebox(0,0)[lb]{\smash{{\SetFigFont{12}{14.4}{\rmdefault}{\mddefault}{\updefault}{\color[rgb]{0,0,0}$B$}%
}}}}
\put(3900,-691){\makebox(0,0)[lb]{\smash{{\SetFigFont{12}{14.4}{\rmdefault}{\mddefault}{\updefault}{\color[rgb]{0,0,0}$A$}%
}}}}
\put(1462,-337){\makebox(0,0)[lb]{\smash{{\SetFigFont{12}{14.4}{\rmdefault}{\mddefault}{\updefault}{\color[rgb]{0,0,0}$C$}%
}}}}
\put(1856, 68){\makebox(0,0)[lb]{\smash{{\SetFigFont{12}{14.4}{\rmdefault}{\mddefault}{\updefault}{\color[rgb]{0,0,0}$C$}%
}}}}
\put(1831,-337){\makebox(0,0)[lb]{\smash{{\SetFigFont{12}{14.4}{\rmdefault}{\mddefault}{\updefault}{\color[rgb]{0,0,0}$B$}%
}}}}
\put(2071,-337){\makebox(0,0)[lb]{\smash{{\SetFigFont{12}{14.4}{\rmdefault}{\mddefault}{\updefault}{\color[rgb]{0,0,0}$A^\prime$}%
}}}}
\put(1146,-691){\makebox(0,0)[lb]{\smash{{\SetFigFont{12}{14.4}{\rmdefault}{\mddefault}{\updefault}{\color[rgb]{0,0,0}$C$}%
}}}}
\put(1344,-691){\makebox(0,0)[lb]{\smash{{\SetFigFont{12}{14.4}{\rmdefault}{\mddefault}{\updefault}{\color[rgb]{0,0,0}$B$}%
}}}}
\put(1738,-691){\makebox(0,0)[lb]{\smash{{\SetFigFont{12}{14.4}{\rmdefault}{\mddefault}{\updefault}{\color[rgb]{0,0,0}$A$}%
}}}}
\put(1500,-691){\makebox(0,0)[lb]{\smash{{\SetFigFont{12}{14.4}{\rmdefault}{\mddefault}{\updefault}{\color[rgb]{0,0,0}$A^\prime$}%
}}}}
\put(1915,-691){\makebox(0,0)[lb]{\smash{{\SetFigFont{12}{14.4}{\rmdefault}{\mddefault}{\updefault}{\color[rgb]{0,0,0}$C$}%
}}}}
\put(2132,-691){\makebox(0,0)[lb]{\smash{{\SetFigFont{12}{14.4}{\rmdefault}{\mddefault}{\updefault}{\color[rgb]{0,0,0}$B$}%
}}}}
\put(2323,-691){\makebox(0,0)[lb]{\smash{{\SetFigFont{12}{14.4}{\rmdefault}{\mddefault}{\updefault}{\color[rgb]{0,0,0}$C$}%
}}}}
\put(1639,464){\makebox(0,0)[lb]{\smash{{\SetFigFont{12}{14.4}{\rmdefault}{\mddefault}{\updefault}{\color[rgb]{0,0,0}$A$}%
}}}}
\put(1375, 68){\makebox(0,0)[lb]{\smash{{\SetFigFont{12}{14.4}{\rmdefault}{\mddefault}{\updefault}{\color[rgb]{0,0,0}$B$}%
}}}}
\put(1108,-337){\makebox(0,0)[lb]{\smash{{\SetFigFont{12}{14.4}{\rmdefault}{\mddefault}{\updefault}{\color[rgb]{0,0,0}$A$}%
}}}}
\put(950,-691){\makebox(0,0)[lb]{\smash{{\SetFigFont{12}{14.4}{\rmdefault}{\mddefault}{\updefault}{\color[rgb]{0,0,0}$B$}%
}}}}
\put(-599,479){\makebox(0,0)[lb]{\smash{{\SetFigFont{7}{8.4}{\rmdefault}{\mddefault}{\updefault}{\color[rgb]{0,0,0}Level $k+1$}%
}}}}
\put(-599, 83){\makebox(0,0)[lb]{\smash{{\SetFigFont{7}{8.4}{\rmdefault}{\mddefault}{\updefault}{\color[rgb]{0,0,0}Level $k$}%
}}}}
\put(-599,-325){\makebox(0,0)[lb]{\smash{{\SetFigFont{7}{8.4}{\rmdefault}{\mddefault}{\updefault}{\color[rgb]{0,0,0}Level $k-1$}%
}}}}
\put(-599,-679){\makebox(0,0)[lb]{\smash{{\SetFigFont{7}{8.4}{\rmdefault}{\mddefault}{\updefault}{\color[rgb]{0,0,0}Level $k-2$}%
}}}}
\put(-599,-1436){\makebox(0,0)[lb]{\smash{{\SetFigFont{7}{8.4}{\rmdefault}{\mddefault}{\updefault}{\color[rgb]{0,0,0}Level $1$}%
}}}}
\end{picture}%

%% file: subtree_0_1_1.pstex_t
\begin{picture}(0,0)%
\includegraphics{subtree_0_1_1.pstex}%
\end{picture}%
\setlength{\unitlength}{3947sp}%
\begingroup\makeatletter\ifx\SetFigFont\undefined%
\gdef\SetFigFont#1#2#3#4#5{%
  \reset@font\fontsize{#1}{#2pt}%
  \fontfamily{#3}\fontseries{#4}\fontshape{#5}%
  \selectfont}%
\fi\endgroup%
\begin{picture}(1556,670)(136,-69)
\put(365,454){\makebox(0,0)[lb]{\smash{{\SetFigFont{12}{14.4}{\rmdefault}{\mddefault}{\updefault}{\color[rgb]{0,0,0}$A$}%
}}}}
\put(151, -5){\makebox(0,0)[lb]{\smash{{\SetFigFont{12}{14.4}{\rmdefault}{\mddefault}{\updefault}{\color[rgb]{0,0,0}$B$}%
}}}}
\put(1435,454){\makebox(0,0)[lb]{\smash{{\SetFigFont{12}{14.4}{\rmdefault}{\mddefault}{\updefault}{\color[rgb]{0,0,0}$A$}%
}}}}
\put(1220, -5){\makebox(0,0)[lb]{\smash{{\SetFigFont{12}{14.4}{\rmdefault}{\mddefault}{\updefault}{\color[rgb]{0,0,0}$B$}%
}}}}
\put(481, -5){\makebox(0,0)[lb]{\smash{{\SetFigFont{12}{14.4}{\rmdefault}{\mddefault}{\updefault}{\color[rgb]{0,0,0}$C$}%
}}}}
\put(1561, -5){\makebox(0,0)[lb]{\smash{{\SetFigFont{12}{14.4}{\rmdefault}{\mddefault}{\updefault}{\color[rgb]{0,0,0}$C$}%
}}}}
\end{picture}%

%% file: subtree_0_1_2.pstex_t
\begin{picture}(0,0)%
\includegraphics{subtree_0_1_2.pstex}%
\end{picture}%
\setlength{\unitlength}{3947sp}%
\begingroup\makeatletter\ifx\SetFigFont\undefined%
\gdef\SetFigFont#1#2#3#4#5{%
  \reset@font\fontsize{#1}{#2pt}%
  \fontfamily{#3}\fontseries{#4}\fontshape{#5}%
  \selectfont}%
\fi\endgroup%
\begin{picture}(2374,1110)(100,-674)
\put(294,-185){\makebox(0,0)[lb]{\smash{{\SetFigFont{12}{14.4}{\rmdefault}{\mddefault}{\updefault}{\color[rgb]{0,0,0}$B$}%
}}}}
\put(736,-185){\makebox(0,0)[lb]{\smash{{\SetFigFont{12}{14.4}{\rmdefault}{\mddefault}{\updefault}{\color[rgb]{0,0,0}$C$}%
}}}}
\put(115,-601){\makebox(0,0)[lb]{\smash{{\SetFigFont{12}{14.4}{\rmdefault}{\mddefault}{\updefault}{\color[rgb]{0,0,0}$A$}%
}}}}
\put(381,-601){\makebox(0,0)[lb]{\smash{{\SetFigFont{12}{14.4}{\rmdefault}{\mddefault}{\updefault}{\color[rgb]{0,0,0}$C$}%
}}}}
\put(919,-601){\makebox(0,0)[lb]{\smash{{\SetFigFont{12}{14.4}{\rmdefault}{\mddefault}{\updefault}{\color[rgb]{0,0,0}$B$}%
}}}}
\put(514,289){\makebox(0,0)[lb]{\smash{{\SetFigFont{12}{14.4}{\rmdefault}{\mddefault}{\updefault}{\color[rgb]{0,0,0}$A$}%
}}}}
\put(613,-601){\makebox(0,0)[lb]{\smash{{\SetFigFont{12}{14.4}{\rmdefault}{\mddefault}{\updefault}{\color[rgb]{0,0,0}$A^\prime$}%
}}}}
\put(1794,-185){\makebox(0,0)[lb]{\smash{{\SetFigFont{12}{14.4}{\rmdefault}{\mddefault}{\updefault}{\color[rgb]{0,0,0}$B$}%
}}}}
\put(2236,-185){\makebox(0,0)[lb]{\smash{{\SetFigFont{12}{14.4}{\rmdefault}{\mddefault}{\updefault}{\color[rgb]{0,0,0}$C$}%
}}}}
\put(1615,-601){\makebox(0,0)[lb]{\smash{{\SetFigFont{12}{14.4}{\rmdefault}{\mddefault}{\updefault}{\color[rgb]{0,0,0}$A$}%
}}}}
\put(1881,-601){\makebox(0,0)[lb]{\smash{{\SetFigFont{12}{14.4}{\rmdefault}{\mddefault}{\updefault}{\color[rgb]{0,0,0}$C$}%
}}}}
\put(2419,-601){\makebox(0,0)[lb]{\smash{{\SetFigFont{12}{14.4}{\rmdefault}{\mddefault}{\updefault}{\color[rgb]{0,0,0}$B$}%
}}}}
\put(2014,289){\makebox(0,0)[lb]{\smash{{\SetFigFont{12}{14.4}{\rmdefault}{\mddefault}{\updefault}{\color[rgb]{0,0,0}$A$}%
}}}}
\put(2113,-601){\makebox(0,0)[lb]{\smash{{\SetFigFont{12}{14.4}{\rmdefault}{\mddefault}{\updefault}{\color[rgb]{0,0,0}$A$}%
}}}}
\end{picture}%

%% file: induction_sub_tree_1.pstex_t
\begin{picture}(0,0)%
\includegraphics{induction_sub_tree_1.pstex}%
\end{picture}%
\setlength{\unitlength}{3947sp}%
\begingroup\makeatletter\ifx\SetFigFont\undefined%
\gdef\SetFigFont#1#2#3#4#5{%
  \reset@font\fontsize{#1}{#2pt}%
  \fontfamily{#3}\fontseries{#4}\fontshape{#5}%
  \selectfont}%
\fi\endgroup%
\begin{picture}(6273,2093)(-164,-1482)
\put(4858,-337){\makebox(0,0)[lb]{\smash{{\SetFigFont{12}{14.4}{\rmdefault}{\mddefault}{\updefault}{\color[rgb]{0,0,0}$A$}%
}}}}
\put(5212,-337){\makebox(0,0)[lb]{\smash{{\SetFigFont{12}{14.4}{\rmdefault}{\mddefault}{\updefault}{\color[rgb]{0,0,0}$C$}%
}}}}
\put(5606, 68){\makebox(0,0)[lb]{\smash{{\SetFigFont{12}{14.4}{\rmdefault}{\mddefault}{\updefault}{\color[rgb]{0,0,0}$C$}%
}}}}
\put(5125, 68){\makebox(0,0)[lb]{\smash{{\SetFigFont{12}{14.4}{\rmdefault}{\mddefault}{\updefault}{\color[rgb]{0,0,0}$B$}%
}}}}
\put(5581,-337){\makebox(0,0)[lb]{\smash{{\SetFigFont{12}{14.4}{\rmdefault}{\mddefault}{\updefault}{\color[rgb]{0,0,0}$B$}%
}}}}
\put(5389,464){\makebox(0,0)[lb]{\smash{{\SetFigFont{12}{14.4}{\rmdefault}{\mddefault}{\updefault}{\color[rgb]{0,0,0}$A$}%
}}}}
\put(5488,-691){\makebox(0,0)[lb]{\smash{{\SetFigFont{12}{14.4}{\rmdefault}{\mddefault}{\updefault}{\color[rgb]{0,0,0}$A$}%
}}}}
\put(5665,-691){\makebox(0,0)[lb]{\smash{{\SetFigFont{12}{14.4}{\rmdefault}{\mddefault}{\updefault}{\color[rgb]{0,0,0}$C$}%
}}}}
\put(5882,-691){\makebox(0,0)[lb]{\smash{{\SetFigFont{12}{14.4}{\rmdefault}{\mddefault}{\updefault}{\color[rgb]{0,0,0}$B$}%
}}}}
\put(6073,-691){\makebox(0,0)[lb]{\smash{{\SetFigFont{12}{14.4}{\rmdefault}{\mddefault}{\updefault}{\color[rgb]{0,0,0}$C$}%
}}}}
\put(5821,-337){\makebox(0,0)[lb]{\smash{{\SetFigFont{12}{14.4}{\rmdefault}{\mddefault}{\updefault}{\color[rgb]{0,0,0}$A$}%
}}}}
\put(5094,-691){\makebox(0,0)[lb]{\smash{{\SetFigFont{12}{14.4}{\rmdefault}{\mddefault}{\updefault}{\color[rgb]{0,0,0}$B$}%
}}}}
\put(4896,-691){\makebox(0,0)[lb]{\smash{{\SetFigFont{12}{14.4}{\rmdefault}{\mddefault}{\updefault}{\color[rgb]{0,0,0}$C$}%
}}}}
\put(4700,-691){\makebox(0,0)[lb]{\smash{{\SetFigFont{12}{14.4}{\rmdefault}{\mddefault}{\updefault}{\color[rgb]{0,0,0}$B$}%
}}}}
\put(5250,-691){\makebox(0,0)[lb]{\smash{{\SetFigFont{12}{14.4}{\rmdefault}{\mddefault}{\updefault}{\color[rgb]{0,0,0}$A$}%
}}}}
\put(1462,-337){\makebox(0,0)[lb]{\smash{{\SetFigFont{12}{14.4}{\rmdefault}{\mddefault}{\updefault}{\color[rgb]{0,0,0}$C$}%
}}}}
\put(1856, 68){\makebox(0,0)[lb]{\smash{{\SetFigFont{12}{14.4}{\rmdefault}{\mddefault}{\updefault}{\color[rgb]{0,0,0}$C$}%
}}}}
\put(1831,-337){\makebox(0,0)[lb]{\smash{{\SetFigFont{12}{14.4}{\rmdefault}{\mddefault}{\updefault}{\color[rgb]{0,0,0}$B$}%
}}}}
\put(2071,-337){\makebox(0,0)[lb]{\smash{{\SetFigFont{12}{14.4}{\rmdefault}{\mddefault}{\updefault}{\color[rgb]{0,0,0}$A^\prime$}%
}}}}
\put(1146,-691){\makebox(0,0)[lb]{\smash{{\SetFigFont{12}{14.4}{\rmdefault}{\mddefault}{\updefault}{\color[rgb]{0,0,0}$C$}%
}}}}
\put(1344,-691){\makebox(0,0)[lb]{\smash{{\SetFigFont{12}{14.4}{\rmdefault}{\mddefault}{\updefault}{\color[rgb]{0,0,0}$B$}%
}}}}
\put(1738,-691){\makebox(0,0)[lb]{\smash{{\SetFigFont{12}{14.4}{\rmdefault}{\mddefault}{\updefault}{\color[rgb]{0,0,0}$A$}%
}}}}
\put(1500,-691){\makebox(0,0)[lb]{\smash{{\SetFigFont{12}{14.4}{\rmdefault}{\mddefault}{\updefault}{\color[rgb]{0,0,0}$A^\prime$}%
}}}}
\put(1915,-691){\makebox(0,0)[lb]{\smash{{\SetFigFont{12}{14.4}{\rmdefault}{\mddefault}{\updefault}{\color[rgb]{0,0,0}$C$}%
}}}}
\put(2132,-691){\makebox(0,0)[lb]{\smash{{\SetFigFont{12}{14.4}{\rmdefault}{\mddefault}{\updefault}{\color[rgb]{0,0,0}$B$}%
}}}}
\put(2323,-691){\makebox(0,0)[lb]{\smash{{\SetFigFont{12}{14.4}{\rmdefault}{\mddefault}{\updefault}{\color[rgb]{0,0,0}$C$}%
}}}}
\put(1639,464){\makebox(0,0)[lb]{\smash{{\SetFigFont{12}{14.4}{\rmdefault}{\mddefault}{\updefault}{\color[rgb]{0,0,0}$A$}%
}}}}
\put(1375, 68){\makebox(0,0)[lb]{\smash{{\SetFigFont{12}{14.4}{\rmdefault}{\mddefault}{\updefault}{\color[rgb]{0,0,0}$B$}%
}}}}
\put(1108,-337){\makebox(0,0)[lb]{\smash{{\SetFigFont{12}{14.4}{\rmdefault}{\mddefault}{\updefault}{\color[rgb]{0,0,0}$A$}%
}}}}
\put(950,-691){\makebox(0,0)[lb]{\smash{{\SetFigFont{12}{14.4}{\rmdefault}{\mddefault}{\updefault}{\color[rgb]{0,0,0}$B$}%
}}}}
\put(-149,479){\makebox(0,0)[lb]{\smash{{\SetFigFont{7}{8.4}{\rmdefault}{\mddefault}{\updefault}{\color[rgb]{0,0,0}Level $k+2$}%
}}}}
\put(-149, 83){\makebox(0,0)[lb]{\smash{{\SetFigFont{7}{8.4}{\rmdefault}{\mddefault}{\updefault}{\color[rgb]{0,0,0}Level $k+1$}%
}}}}
\put(-149,-325){\makebox(0,0)[lb]{\smash{{\SetFigFont{7}{8.4}{\rmdefault}{\mddefault}{\updefault}{\color[rgb]{0,0,0}Level $k$}%
}}}}
\put(-149,-679){\makebox(0,0)[lb]{\smash{{\SetFigFont{7}{8.4}{\rmdefault}{\mddefault}{\updefault}{\color[rgb]{0,0,0}Level $k-1$}%
}}}}
\put(-149,-1436){\makebox(0,0)[lb]{\smash{{\SetFigFont{7}{8.4}{\rmdefault}{\mddefault}{\updefault}{\color[rgb]{0,0,0}Level $1$}%
}}}}
\put(3676,479){\makebox(0,0)[lb]{\smash{{\SetFigFont{7}{8.4}{\rmdefault}{\mddefault}{\updefault}{\color[rgb]{0,0,0}Level $k+2$}%
}}}}
\put(3676, 83){\makebox(0,0)[lb]{\smash{{\SetFigFont{7}{8.4}{\rmdefault}{\mddefault}{\updefault}{\color[rgb]{0,0,0}Level $k+1$}%
}}}}
\put(3676,-325){\makebox(0,0)[lb]{\smash{{\SetFigFont{7}{8.4}{\rmdefault}{\mddefault}{\updefault}{\color[rgb]{0,0,0}Level $k$}%
}}}}
\put(3676,-679){\makebox(0,0)[lb]{\smash{{\SetFigFont{7}{8.4}{\rmdefault}{\mddefault}{\updefault}{\color[rgb]{0,0,0}Level $k-1$}%
}}}}
\put(3676,-1436){\makebox(0,0)[lb]{\smash{{\SetFigFont{7}{8.4}{\rmdefault}{\mddefault}{\updefault}{\color[rgb]{0,0,0}Level $1$}%
}}}}
\end{picture}%

%% file: subtree_0_2_1.pstex_t
\begin{picture}(0,0)%
\includegraphics{subtree_0_2_1.pstex}%
\end{picture}%
\setlength{\unitlength}{3947sp}%
\begingroup\makeatletter\ifx\SetFigFont\undefined%
\gdef\SetFigFont#1#2#3#4#5{%
  \reset@font\fontsize{#1}{#2pt}%
  \fontfamily{#3}\fontseries{#4}\fontshape{#5}%
  \selectfont}%
\fi\endgroup%
\begin{picture}(1586,670)(106,-69)
\put(365,454){\makebox(0,0)[lb]{\smash{{\SetFigFont{12}{14.4}{\rmdefault}{\mddefault}{\updefault}{\color[rgb]{0,0,0}$B$}%
}}}}
\put(1435,454){\makebox(0,0)[lb]{\smash{{\SetFigFont{12}{14.4}{\rmdefault}{\mddefault}{\updefault}{\color[rgb]{0,0,0}$B$}%
}}}}
\put(121, -5){\makebox(0,0)[lb]{\smash{{\SetFigFont{12}{14.4}{\rmdefault}{\mddefault}{\updefault}{\color[rgb]{0,0,0}$A$}%
}}}}
\put(565, -5){\makebox(0,0)[lb]{\smash{{\SetFigFont{12}{14.4}{\rmdefault}{\mddefault}{\updefault}{\color[rgb]{0,0,0}$C$}%
}}}}
\put(1201, -5){\makebox(0,0)[lb]{\smash{{\SetFigFont{12}{14.4}{\rmdefault}{\mddefault}{\updefault}{\color[rgb]{0,0,0}$A$}%
}}}}
\put(1645, -5){\makebox(0,0)[lb]{\smash{{\SetFigFont{12}{14.4}{\rmdefault}{\mddefault}{\updefault}{\color[rgb]{0,0,0}$C$}%
}}}}
\end{picture}%

%% file: subtree_0_2_2.pstex_t
\begin{picture}(0,0)%
\includegraphics{subtree_0_2_2.pstex}%
\end{picture}%
\setlength{\unitlength}{3947sp}%
\begingroup\makeatletter\ifx\SetFigFont\undefined%
\gdef\SetFigFont#1#2#3#4#5{%
  \reset@font\fontsize{#1}{#2pt}%
  \fontfamily{#3}\fontseries{#4}\fontshape{#5}%
  \selectfont}%
\fi\endgroup%
\begin{picture}(2374,1110)(100,-674)
\put(514,289){\makebox(0,0)[lb]{\smash{{\SetFigFont{12}{14.4}{\rmdefault}{\mddefault}{\updefault}{\color[rgb]{0,0,0}$B$}%
}}}}
\put(294,-185){\makebox(0,0)[lb]{\smash{{\SetFigFont{12}{14.4}{\rmdefault}{\mddefault}{\updefault}{\color[rgb]{0,0,0}$A$}%
}}}}
\put(736,-185){\makebox(0,0)[lb]{\smash{{\SetFigFont{12}{14.4}{\rmdefault}{\mddefault}{\updefault}{\color[rgb]{0,0,0}$C$}%
}}}}
\put(115,-601){\makebox(0,0)[lb]{\smash{{\SetFigFont{12}{14.4}{\rmdefault}{\mddefault}{\updefault}{\color[rgb]{0,0,0}$B$}%
}}}}
\put(381,-601){\makebox(0,0)[lb]{\smash{{\SetFigFont{12}{14.4}{\rmdefault}{\mddefault}{\updefault}{\color[rgb]{0,0,0}$C$}%
}}}}
\put(2014,289){\makebox(0,0)[lb]{\smash{{\SetFigFont{12}{14.4}{\rmdefault}{\mddefault}{\updefault}{\color[rgb]{0,0,0}$B$}%
}}}}
\put(1794,-185){\makebox(0,0)[lb]{\smash{{\SetFigFont{12}{14.4}{\rmdefault}{\mddefault}{\updefault}{\color[rgb]{0,0,0}$A$}%
}}}}
\put(2236,-185){\makebox(0,0)[lb]{\smash{{\SetFigFont{12}{14.4}{\rmdefault}{\mddefault}{\updefault}{\color[rgb]{0,0,0}$C$}%
}}}}
\put(1615,-601){\makebox(0,0)[lb]{\smash{{\SetFigFont{12}{14.4}{\rmdefault}{\mddefault}{\updefault}{\color[rgb]{0,0,0}$B$}%
}}}}
\put(1881,-601){\makebox(0,0)[lb]{\smash{{\SetFigFont{12}{14.4}{\rmdefault}{\mddefault}{\updefault}{\color[rgb]{0,0,0}$C$}%
}}}}
\put(2125,-601){\makebox(0,0)[lb]{\smash{{\SetFigFont{12}{14.4}{\rmdefault}{\mddefault}{\updefault}{\color[rgb]{0,0,0}$A$}%
}}}}
\put(625,-601){\makebox(0,0)[lb]{\smash{{\SetFigFont{12}{14.4}{\rmdefault}{\mddefault}{\updefault}{\color[rgb]{0,0,0}$A^\prime$}%
}}}}
\put(889,-601){\makebox(0,0)[lb]{\smash{{\SetFigFont{12}{14.4}{\rmdefault}{\mddefault}{\updefault}{\color[rgb]{0,0,0}$B$}%
}}}}
\put(2389,-601){\makebox(0,0)[lb]{\smash{{\SetFigFont{12}{14.4}{\rmdefault}{\mddefault}{\updefault}{\color[rgb]{0,0,0}$B$}%
}}}}
\end{picture}%

%% file: induction_sub_tree_2.pstex_t
\begin{picture}(0,0)%
\includegraphics{induction_sub_tree_2.pstex}%
\end{picture}%
\setlength{\unitlength}{3947sp}%
\begingroup\makeatletter\ifx\SetFigFont\undefined%
\gdef\SetFigFont#1#2#3#4#5{%
  \reset@font\fontsize{#1}{#2pt}%
  \fontfamily{#3}\fontseries{#4}\fontshape{#5}%
  \selectfont}%
\fi\endgroup%
\begin{picture}(6273,2093)(-164,-1482)
\put(1462,-337){\makebox(0,0)[lb]{\smash{{\SetFigFont{12}{14.4}{\rmdefault}{\mddefault}{\updefault}{\color[rgb]{0,0,0}$C$}%
}}}}
\put(1856, 68){\makebox(0,0)[lb]{\smash{{\SetFigFont{12}{14.4}{\rmdefault}{\mddefault}{\updefault}{\color[rgb]{0,0,0}$C$}%
}}}}
\put(1831,-337){\makebox(0,0)[lb]{\smash{{\SetFigFont{12}{14.4}{\rmdefault}{\mddefault}{\updefault}{\color[rgb]{0,0,0}$B$}%
}}}}
\put(2071,-337){\makebox(0,0)[lb]{\smash{{\SetFigFont{12}{14.4}{\rmdefault}{\mddefault}{\updefault}{\color[rgb]{0,0,0}$A^\prime$}%
}}}}
\put(1738,-691){\makebox(0,0)[lb]{\smash{{\SetFigFont{12}{14.4}{\rmdefault}{\mddefault}{\updefault}{\color[rgb]{0,0,0}$A$}%
}}}}
\put(1915,-691){\makebox(0,0)[lb]{\smash{{\SetFigFont{12}{14.4}{\rmdefault}{\mddefault}{\updefault}{\color[rgb]{0,0,0}$C$}%
}}}}
\put(2132,-691){\makebox(0,0)[lb]{\smash{{\SetFigFont{12}{14.4}{\rmdefault}{\mddefault}{\updefault}{\color[rgb]{0,0,0}$B$}%
}}}}
\put(2323,-691){\makebox(0,0)[lb]{\smash{{\SetFigFont{12}{14.4}{\rmdefault}{\mddefault}{\updefault}{\color[rgb]{0,0,0}$C$}%
}}}}
\put(-149,479){\makebox(0,0)[lb]{\smash{{\SetFigFont{7}{8.4}{\rmdefault}{\mddefault}{\updefault}{\color[rgb]{0,0,0}Level $k+2$}%
}}}}
\put(-149, 83){\makebox(0,0)[lb]{\smash{{\SetFigFont{7}{8.4}{\rmdefault}{\mddefault}{\updefault}{\color[rgb]{0,0,0}Level $k+1$}%
}}}}
\put(-149,-325){\makebox(0,0)[lb]{\smash{{\SetFigFont{7}{8.4}{\rmdefault}{\mddefault}{\updefault}{\color[rgb]{0,0,0}Level $k$}%
}}}}
\put(-149,-679){\makebox(0,0)[lb]{\smash{{\SetFigFont{7}{8.4}{\rmdefault}{\mddefault}{\updefault}{\color[rgb]{0,0,0}Level $k-1$}%
}}}}
\put(-149,-1436){\makebox(0,0)[lb]{\smash{{\SetFigFont{7}{8.4}{\rmdefault}{\mddefault}{\updefault}{\color[rgb]{0,0,0}Level $1$}%
}}}}
\put(3676,479){\makebox(0,0)[lb]{\smash{{\SetFigFont{7}{8.4}{\rmdefault}{\mddefault}{\updefault}{\color[rgb]{0,0,0}Level $k+2$}%
}}}}
\put(3676, 83){\makebox(0,0)[lb]{\smash{{\SetFigFont{7}{8.4}{\rmdefault}{\mddefault}{\updefault}{\color[rgb]{0,0,0}Level $k+1$}%
}}}}
\put(3676,-325){\makebox(0,0)[lb]{\smash{{\SetFigFont{7}{8.4}{\rmdefault}{\mddefault}{\updefault}{\color[rgb]{0,0,0}Level $k$}%
}}}}
\put(3676,-679){\makebox(0,0)[lb]{\smash{{\SetFigFont{7}{8.4}{\rmdefault}{\mddefault}{\updefault}{\color[rgb]{0,0,0}Level $k-1$}%
}}}}
\put(3676,-1436){\makebox(0,0)[lb]{\smash{{\SetFigFont{7}{8.4}{\rmdefault}{\mddefault}{\updefault}{\color[rgb]{0,0,0}Level $1$}%
}}}}
\put(1639,464){\makebox(0,0)[lb]{\smash{{\SetFigFont{12}{14.4}{\rmdefault}{\mddefault}{\updefault}{\color[rgb]{0,0,0}$B$}%
}}}}
\put(1375, 68){\makebox(0,0)[lb]{\smash{{\SetFigFont{12}{14.4}{\rmdefault}{\mddefault}{\updefault}{\color[rgb]{0,0,0}$A$}%
}}}}
\put(1108,-337){\makebox(0,0)[lb]{\smash{{\SetFigFont{12}{14.4}{\rmdefault}{\mddefault}{\updefault}{\color[rgb]{0,0,0}$B$}%
}}}}
\put(950,-691){\makebox(0,0)[lb]{\smash{{\SetFigFont{12}{14.4}{\rmdefault}{\mddefault}{\updefault}{\color[rgb]{0,0,0}$A$}%
}}}}
\put(1146,-691){\makebox(0,0)[lb]{\smash{{\SetFigFont{12}{14.4}{\rmdefault}{\mddefault}{\updefault}{\color[rgb]{0,0,0}$C$}%
}}}}
\put(1344,-691){\makebox(0,0)[lb]{\smash{{\SetFigFont{12}{14.4}{\rmdefault}{\mddefault}{\updefault}{\color[rgb]{0,0,0}$B$}%
}}}}
\put(1500,-691){\makebox(0,0)[lb]{\smash{{\SetFigFont{12}{14.4}{\rmdefault}{\mddefault}{\updefault}{\color[rgb]{0,0,0}$A^\prime$}%
}}}}
\put(5212,-337){\makebox(0,0)[lb]{\smash{{\SetFigFont{12}{14.4}{\rmdefault}{\mddefault}{\updefault}{\color[rgb]{0,0,0}$C$}%
}}}}
\put(5606, 68){\makebox(0,0)[lb]{\smash{{\SetFigFont{12}{14.4}{\rmdefault}{\mddefault}{\updefault}{\color[rgb]{0,0,0}$C$}%
}}}}
\put(5581,-337){\makebox(0,0)[lb]{\smash{{\SetFigFont{12}{14.4}{\rmdefault}{\mddefault}{\updefault}{\color[rgb]{0,0,0}$B$}%
}}}}
\put(5488,-691){\makebox(0,0)[lb]{\smash{{\SetFigFont{12}{14.4}{\rmdefault}{\mddefault}{\updefault}{\color[rgb]{0,0,0}$A$}%
}}}}
\put(5665,-691){\makebox(0,0)[lb]{\smash{{\SetFigFont{12}{14.4}{\rmdefault}{\mddefault}{\updefault}{\color[rgb]{0,0,0}$C$}%
}}}}
\put(5882,-691){\makebox(0,0)[lb]{\smash{{\SetFigFont{12}{14.4}{\rmdefault}{\mddefault}{\updefault}{\color[rgb]{0,0,0}$B$}%
}}}}
\put(6073,-691){\makebox(0,0)[lb]{\smash{{\SetFigFont{12}{14.4}{\rmdefault}{\mddefault}{\updefault}{\color[rgb]{0,0,0}$C$}%
}}}}
\put(5821,-337){\makebox(0,0)[lb]{\smash{{\SetFigFont{12}{14.4}{\rmdefault}{\mddefault}{\updefault}{\color[rgb]{0,0,0}$A$}%
}}}}
\put(5094,-691){\makebox(0,0)[lb]{\smash{{\SetFigFont{12}{14.4}{\rmdefault}{\mddefault}{\updefault}{\color[rgb]{0,0,0}$B$}%
}}}}
\put(4896,-691){\makebox(0,0)[lb]{\smash{{\SetFigFont{12}{14.4}{\rmdefault}{\mddefault}{\updefault}{\color[rgb]{0,0,0}$C$}%
}}}}
\put(5250,-691){\makebox(0,0)[lb]{\smash{{\SetFigFont{12}{14.4}{\rmdefault}{\mddefault}{\updefault}{\color[rgb]{0,0,0}$A$}%
}}}}
\put(5389,464){\makebox(0,0)[lb]{\smash{{\SetFigFont{12}{14.4}{\rmdefault}{\mddefault}{\updefault}{\color[rgb]{0,0,0}$B$}%
}}}}
\put(5125, 68){\makebox(0,0)[lb]{\smash{{\SetFigFont{12}{14.4}{\rmdefault}{\mddefault}{\updefault}{\color[rgb]{0,0,0}$A$}%
}}}}
\put(4858,-337){\makebox(0,0)[lb]{\smash{{\SetFigFont{12}{14.4}{\rmdefault}{\mddefault}{\updefault}{\color[rgb]{0,0,0}$B$}%
}}}}
\put(4700,-691){\makebox(0,0)[lb]{\smash{{\SetFigFont{12}{14.4}{\rmdefault}{\mddefault}{\updefault}{\color[rgb]{0,0,0}$A$}%
}}}}
\end{picture}%